\documentclass[12pt]{article}
\usepackage{amsmath}
\usepackage[dvips]{graphicx}
\usepackage[nosort]{cite}

\begin{document}

\baselineskip=17pt

\begin{titlepage}
\rightline{\tt arXiv:1108.5317}
\rightline{\tt UT-Komaba/11-6}
\begin{center}
\vskip 3.5cm
{\Large \bf {Solutions from boundary condition changing operators}}\\
\vskip 0.4cm
{\Large \bf {in open superstring field theory}}
\vskip 1.0cm
{\large {Toshifumi Noumi and Yuji Okawa}}
\vskip 1.0cm
{\it {Institute of Physics, University of Tokyo}}\\
{\it {Komaba, Meguro-ku, Tokyo 153-8902, Japan}}\\
tnoumi@hep1.c.u-tokyo.ac.jp, okawa@hep1.c.u-tokyo.ac.jp

\vskip 3.0cm

{\bf Abstract}
\end{center}

\noindent
We construct analytic solutions of open superstring field theory
in the Berkovits formulation
using boundary condition changing operators
under some regularity conditions,
extending the previous construction in the bosonic string.
We also consider the gauge-invariant observables
corresponding to closed string one-point functions on the disk.
We analytically calculate the gauge-invariant observables for the solutions
both in the bosonic string and in the superstring
and find the expected change of boundary conditions of the disk.
\end{titlepage}

\newpage

\section{Introduction}
\setcounter{equation}{0}

String field theory can be thought of as a universal, effective theory
when elementary excitations are string-like.
To eliminate ghosts from such string-like excitations
in unphysical directions,
any covariant description in terms of a spacetime field theory
would require gauge invariance.
Then the gauge invariance seems to determine
the interacting theory uniquely.
This has been a guiding principle in constructing covariant string field theory~\cite{Witten:1985cc,Zwiebach:1992ie}.

\medskip
On the other hand, we know from the perturbative world-sheet formulation
of string theory that consistent backgrounds are described
by conformal field theories in two dimensions.
The classical equation of motion of string field theory
determined by the spacetime gauge invariance
should therefore reproduce this requirement of conformal invariance
in the world-sheet perspective.
We hope that deeper understanding of the relation
between the world-sheet conformal invariance
and the spacetime gauge invariance
would reveal aspects of the non-perturbative theory
behind the perturbative string theory.

\medskip
In the case of the open string, a consistent background
is given by a choice of boundary conformal field theory (BCFT),
and different open string backgrounds correspond
to different conformal boundary conditions.
We would therefore like to have a systematic construction
of solutions in open string field theory
for a given BCFT.
Since the change of boundary conditions can be described
by insertions of boundary condition changing
 (bcc) operators in the original BCFT,
one possible approach to such systematic constructions
would be to use bcc operators.

\medskip
Since the construction of an analytic solution
for tachyon condensation by Schnabl~\cite{Schnabl:2005gv},
an impressive amount of analytic results for open string field theory
have been obtained~\cite{Okawa:2006vm, Fuchs:2006hw, Fuchs:2006an,
Rastelli:2006ap, Ellwood:2006ba,Fuji:2006me,
Fuchs:2006gs, Okawa:2006sn, Erler:2006hw, Erler:2006ww, Schnabl:2007az,
Kiermaier:2007ba, Erler:2007rh,
Okawa:2007ri, Fuchs:2007yy, Okawa:2007it,
Ellwood:2007xr,Kishimoto:2007bb, Fuchs:2007gw,
Kiermaier:2007vu, Erler:2007xt, Rastelli:2007gg, Kiermaier:2007ki,
Kwon:2007mh,
Takahashi:2007du,Kiermaier:2007jg,Kwon:2008ap,Hellerman:2008wp,Ellwood:2008jh,Kawano:2008ry,
Aref'eva:2008ad,
Ishida:2008jc,Kawano:2008jv,Kiermaier:2008jy,Fuchs:2008zx,Fuchs:2008cc,Asano:2008iu,Kishimoto:2008zj,Kiermaier:2008qu,Barnaby:2008pt,
Aref'eva:2009ac, Kishimoto:2009cz,Ellwood:2009zf,Aref'eva:2009sj,
Arroyo:2009ec,
Kroyter:2009bg,Erler:2009uj,AldoArroyo:2009hf, Beaujean:2009rb,
Arroyo:2010fq, Zeze:2010jv, Schnabl:2010tb, Arefeva:2010yd,
Zeze:2010sr, Arroyo:2010sy, Erler:2010pr, Bonora:2010hi,
Kiermaier:2010cf,Murata:2011ex,Arroyo:2011zt,Bonora:2011ri,Erler:2011tc,Bonora:2011ru,Inatomi:2011xr}.
In~\cite{Kiermaier:2010cf}, analytic solutions of open bosonic string field theory
were constructed using bcc operators
when they satisfy some regularity conditions.
The starting point of the construction was
the solutions in Schnabl gauge for marginal deformations
when operator products of the marginal operators are
regular~\cite{Schnabl:2007az,Kiermaier:2007ba}.
The solutions take the form of a superposition of wedge-based states.\footnote{
We denote wedge states~\cite{Rastelli:2000iu,Schnabl:2002gg} with operator insertions by wedge-based states.
}
When we write the wedge state $W_\alpha$ as $W_\alpha = e^{\alpha K}$,\footnote{
Products of string fields in this paper are defined
using Witten's star product~\cite{Witten:1985cc}.
}
the solutions can be written in terms of the states $K$, $B$, $c$, and $V$,
where these states are based on the wedge state  $W_0$ of zero width
with a line integral of the energy-momentum tensor and the $b$ ghost for $K$ and $B$, respectively,
and with a local insertion of the $c$ ghost and the marginal operator for $c$ and $V$, respectively.\footnote{
We follow the conventions of~\cite{Okawa:2006vm},
but the states
are rescaled as
 $K_{\rm here} = (\pi/2) \, K_{\rm there}$,
 $B_{\rm here} = (\pi/2) \, B_{\rm there}$,
 and
 $c_{\rm here} = (2/\pi) \, c_{\rm there}$.}
The solutions in~\cite{Schnabl:2007az,Kiermaier:2007ba} were later generalized by Erler~\cite{Erler:2007rh}.
Just as the tachyon vacuum solution by Schnabl~\cite{Schnabl:2005gv} written in terms of 
$K$, $B$, and $c$
was generalized by replacing $e^{K/2}$ with an arbitrary function $f(K)$ of $K$ in~\cite{Okawa:2006vm},
the solutions in~\cite{Schnabl:2007az,Kiermaier:2007ba} were generalized
by replacing $e^{K/2}$ with an arbitrary function $f(K)$ of $K$.
The resulting solutions depend explicitly on $V$
and thus cannot in general be written in terms of bcc operators.
In~\cite{Kiermaier:2010cf} it was found
that the solutions can be written in terms of bcc operators
for a special choice\footnote{
Incidentally, this is the same choice of $f(K)$ in the phantomless solution
for tachyon condensation~\cite{Erler:2009uj} by Erler and Schnabl.
}
 of the function $f(K)$ given by $f(K) = 1/\sqrt{1-K}$.
In this case the solutions depend on $V$ only through bcc operators,
and
one can show that the solutions satisfy the equation of motion
when bcc operators obey some regularity conditions.
Another important feature of the solutions using bcc operators
is that one can construct the solutions only from the information
on three-point functions on a disk of a pair of bcc operators
and an arbitrary operator in the original BCFT,
while the solutions in Schnabl gauge~\cite{Schnabl:2007az,Kiermaier:2007ba}
depend in a complicated way
on multi-point functions
of the marginal operator.

\medskip
The purpose of this paper is to extend the construction in~\cite{Kiermaier:2010cf}
to open superstring field theory in the Berkovits formulation\cite{Berkovits:1995ab}.
Namely, we would like to construct solutions of open superstring field theory
written in terms of bcc operators.
Actually, this can be immediately achieved
by combining the following two observations.
The first one is that solutions in~\cite{Kiermaier:2010cf}
can be written only in terms of
bcc operators, the energy-momentum tensor, the $b$ ghost,
and the BRST operator $Q$
without explicitly using the $c$ ghost.
We can prove that the solutions satisfy the equation of motion
\begin{equation}
\label{bosonic-equation}
Q \Psi + \Psi^2 = 0
\end{equation}
of open bosonic string field theory, where $\Psi$ is the open string field of ghost number $1$,
only using the relations of these operators which we call $KB \sigma$ algebra.
Since this $KB \sigma$ algebra also holds in open superstring field theory,
we can construct a state $\Psi$ in open superstring field theory
satisfying the bosonic equation of motion~(\ref{bosonic-equation}).
It is important that the $c$ ghost does not appear explicitly
in this form of the solutions because the BRST transformation
of the $c$ ghost takes a different form in the superstring
and requires an extension of the algebra
to the superconformal ghost sector.

\medskip
The second observation is that
one can construct solutions of open superstring field theory from
a state $\Psi$ in the small Hilbert space
satisfying the bosonic string equation of motion~(\ref{bosonic-equation})
by inserting an operator $R (t)$ satisfying $Q \cdot R (t) = 1$ appropriately\cite{Fuchs:2008zx,Erler:2010pr}. 
We can therefore construct solutions in open superstring field theory
written in terms of bcc operators
simply by inserting $R (t)$ appropriately to the state $\Psi$
in the $KB \sigma$ algebra.
This is the first main result of this paper.

\medskip
Since the solutions are written in terms of bcc operators,
we expect that they describe the BCFT
associated with the bcc operators we chose.
As we mentioned earlier,
the solutions in~\cite{Kiermaier:2010cf} were originally constructed
by rewriting solutions for marginal deformations with regular operator products,
and thus these solutions in the bosonic string do describe the BCFT
deformed by regular marginal deformations.
We can also relate
our solutions in the superstring 
to known solutions for regular marginal deformations.
In this paper, we provide more direct evidence of the correspondence
between the solutions using bcc operators and the associated BCFT.
It was conjectured by Ellwood~\cite{Ellwood:2008jh}
that the gauge-invariant observables~\cite{Hashimoto:2001sm,Gaiotto:2001ji}
evaluated for a solution
correspond to closed string one-point functions
on the disk 
with the modified boundary conditions associated with the solution.
We analytically calculate the gauge-invariant observables
for the solutions using bcc operators both in the bosonic string
and in the superstring and find the conjectured change
of boundary conditions of the disk.
This is the second main result of this paper.

\medskip
The organization of this paper is as follows.
In section~\ref{section_bosonic}
we review the solutions~\cite{Kiermaier:2010cf} in the bosonic string.
We introduce the $KB\sigma$ algebra
and demonstrate that the equation of motion is satisfied
only using the $KB\sigma$ algebra.
In section~\ref{section_super}
we construct solutions in the superstring.
After discussing the general prescription
for the construction of solutions in the Berkovits formulation
of open superstring field theory,
we apply it to the solutions using bcc operators.
In section~\ref{section_gauge-invariant}
we analytically calculate the gauge-invariant observables
for the solutions using bcc operators both in the bosonic string
and in the superstring and find the conjectured change
of boundary conditions of the disk.
Section~\ref{discussion} is devoted to discussion.

\section{Solutions to the bosonic string equation of motion}
\label{section_bosonic}
\setcounter{equation}{0}
\subsection{$KB\sigma$ algebra}
\label{subsection_KBsigma}
\setcounter{equation}{0}
The wedge state $W_\alpha$ with $\alpha \ge 0$ is defined by
its BPZ inner product $\langle \, \varphi, W_\alpha \, \rangle$ as follows:
\begin{equation}
\label{wedge-state}
\langle \, \varphi,W_\alpha \, \rangle
=\left\langle \, f\circ \varphi(0) \, \right\rangle_{C_{\alpha+1}}.
\end{equation}
Here and in what follows we denote
a generic state in the Fock space by $\varphi$
and its corresponding operator in the state-operator mapping by $\varphi (\xi)$.
We denote the conformal transformation of $\varphi (\xi)$
under the map $f(\xi)$ by $f \circ \varphi (\xi)$, where
\begin{equation}
f(\xi) = \frac{2}{\pi} \, \arctan \xi \,.
\end{equation}
The coordinate $z$ related through $z = f(\xi)$
to the coordinate $\xi$
on the upper half-plane used in the standard state-operator mapping
is called the sliver frame.
The correlation function is evaluated on the surface $C_{\alpha+1}$,
which is the semi-infinite strip obtained from the upper half-plane of $z$
by the identification $z \sim z+\alpha+1$.
We usually use the region $-1/2 \le \Re z \le 1/2+\alpha$ for $C_{\alpha+1}$.

\medskip
Just as the line integral $L_0$ of the energy-momentum tensor
generates a surface $e^{-t L_0}$ in the standard open string strip coordinate,
the wedge state $W_\alpha$ can be thought of as being generated
by a line integral of the energy-momentum tensor in the sliver frame.
We denote the wedge state $W_0$ of zero width
with an insertion of the line integral by $K$
and write the wedge state $W_\alpha$ as
\begin{equation}
W_\alpha = e^{\alpha K} \,.
\end{equation}
An explicit definition of the state $K$ is given by
\begin{equation}
\left\langle\varphi,K\right\rangle
=\biggl\langle \, f\circ \varphi(0)\int_{\frac{1}{2}+i\infty}^{\frac{1}{2}-i\infty}\frac{dz}{2\pi i} \, T(z) \, \biggr\rangle_{C_{1}},
\end{equation}
where $T(z)$ is the energy-momentum tensor
and we use the doubling trick.
Note that the line integral is from a boundary to the open string mid-point
before using the doubling trick, while the line integral $L_0$ is from a boundary
to the other boundary.

\medskip
Just as the line integral $L_0$ is the BRST transformation
of the line integral $b_0$ of the $b$ ghost,
the line integral that generates the wedge state
is the BRST transformation of the same line integral
with the energy-momentum tensor replaced by the $b$ ghost.
Correspondingly, we define the state $B$ by
\begin{equation}
\langle \, \varphi,B \, \rangle
=\biggl\langle \, f\circ \varphi(0)\int_{\frac{1}{2}+i\infty}^{\frac{1}{2}-i\infty}\frac{dz}{2\pi i} \, b(z) \, \biggr\rangle_{C_{1}}.
\end{equation}
By construction, the state $K$ is the BRST transformation of $B$.
Another important property of the state $B$ is that $B^2 = 0$.

\medskip
Let us next consider wedge states with modified boundary conditions.
The change of boundary conditions on a segment
of the world-sheet boundary from a point $a$ to a point $b$
can be described by inserting a pair of bcc operators
$\sigma_L (a)$ and $\sigma_R (b)$.
We denote the wedge state $W_\alpha = e^{\alpha K}$
with modified boundary conditions by $\sigma_L \, e^{\alpha K} \sigma_R$:
\begin{equation}
\langle \, \varphi, \sigma_L  \, e^{\alpha K} \sigma_R \, \rangle
=\left\langle f\circ \varphi(0) \,
\sigma_L\left(\tfrac{1}{2}\right)\sigma_R\left(\alpha+\tfrac{1}{2}\right)\right\rangle_{C_{\alpha+1}}.
\end{equation}
In other words, we denote the state based on the wedge state $W_0$ of zero width
with an insertion of $\sigma_L (t)$ by $\sigma_L$
and $W_0$ with an insertion of $\sigma_R (t)$ by $\sigma_R$.
Since bcc operators are in the matter sector, they commute
with any line integral of the $b$ ghost.
We therefore have $[ B\, , \sigma_L \, ] = 0$ and $[ B\, , \sigma_R \, ] = 0$.

\medskip
In~\cite{Kiermaier:2010cf} solutions written in terms of $K$, $B$, $\sigma_L$, 
and $\sigma_R$
were constructed when operator products of bcc operators are regular,
while in general they have singular operator products.
The regularity conditions on bcc operators can be stated as
\begin{equation}
\label{conditions27}
\lim_{\epsilon\rightarrow+0}\sigma_L(t) \, \sigma_R(t+\epsilon)=1 \,,\hspace{3mm}
\lim_{\epsilon\rightarrow+0}\sigma_L(a) \, \sigma_R(b) \, \sigma_L(b+\epsilon) \, \sigma_R(c)
=\sigma_L(a) \, \sigma_R(c) \,,
\end{equation}
and in the language of the states $\sigma_L$ and $\sigma_R$
the conditions are translated into the relations
\begin{equation}
\sigma_L \, \sigma_R = \sigma_R \, \sigma_L = 1 \,.
\end{equation}
When we prove that the equation of motion is satisfied,
we use states such as $\sigma_L K\sigma_R$,
and it was implicitly assumed in~\cite{Kiermaier:2010cf}
that these states are well defined.
However, if the operator product of $\sigma_L(t)$ and $\sigma_R(t)$,
for example,
takes
the form $\sigma_L(0)\,\sigma_R(t)=1+t^\alpha\mathcal{A}\,(0)+\ldots$ with $0<\alpha<1$,
the state $\sigma_L K\sigma_R$ is singular.\footnote{
We would like to thank the referee of this paper for pointing this out.
}
It is also possible that
operator products of more bcc operators develop similar singularities.
In this paper we assume that there are no such singularities in the matter sector.

\medskip
To summarize, we have defined the states $K$, $B$, $\sigma_L$, and $\sigma_R$
satisfying the relations
\begin{equation}
\label{KBsigma_1}
B^2 = 0 \,, \quad
[ \, B,\sigma_L \, ] = 0 \,, \quad [ \, B,\sigma_R \, ] = 0 \,, \quad 
\sigma_L \, \sigma_R = 1 \,, \quad  \sigma_R \, \sigma_L=1 \,, 
\end{equation}
and $K=QB$.
One can show that
the solutions in~\cite{Kiermaier:2010cf} satisfy the equation of motion~(\ref{bosonic-equation})
only from these relations.
While we have given an explicit realization of these states,
one can construct a solution from
any Grassmann-odd state $B$
and any Grassmann-even states $\sigma_L$ and $\sigma_R$
satisfying the relations~(\ref{KBsigma_1}),
and $K=QB$ can be regarded as a definition of $K$.

\medskip
By considering the BRST transformation of each of the relations in~(\ref{KBsigma_1}),
we find
\begin{equation}
\label{KBsigma_2}
\begin{split}
& [ \, K, B \, ] = 0 \,,~~
\{ \, B,Q \sigma_L \, \} = [ \, K,\sigma_L \, ] = -[ \, 1-K, \sigma_L \, ] \,,~~
\{ \, B,Q \sigma_R \, \} = [ \, K,\sigma_R \, ] = -[ \, 1-K,\sigma_R \, ] \,,\\
& \left(Q \sigma_L\right) \, \sigma_R + \sigma_L \, \left(Q \sigma_R\right) = 0 \,, \quad 
\left(Q \sigma_R\right) \, \sigma_L + \sigma_R \, \left(Q \sigma_L\right) = 0 \,.
\end{split}
\end{equation}
We use these relations as well as~(\ref{KBsigma_1}).

\subsection{Solutions using bcc operators in the bosonic string}
The solution $\Psi$ written in terms of $K$, $B$, $\sigma_L$, and $\sigma_R$
is given by~\cite{Kiermaier:2010cf}
\begin{equation}
\label{bosonic_solutions_with_bcc}
\Psi=-\frac{1}{\sqrt{1-K}}\left(Q \sigma_L\right) \sigma_R \, \frac{1}{\sqrt{1-K}}
-\frac{1}{\sqrt{1-K}}\left(Q \sigma_L\right)\frac{B}{1-K}\left(Q \sigma_R\right)\frac{1}{\sqrt{1-K}} \,,
\end{equation} 
where
\begin{equation}
\label{laplace-transf}
\frac{1}{1-K} = \int_0^\infty dt \, e^{-t} \, e^{t K} \,, \qquad
\frac{1}{\sqrt{1-K}} = \int_0^\infty dt \, \frac{e^{-t}}{\sqrt{\pi t}} \, e^{t K} \,.
\end{equation}
It is convenient to introduce a non-real solution $\widetilde{\Psi}$
related to $\Psi$ as
\begin{equation}
\label{tilde-Psi}
\begin{split}
\widetilde{\Psi}&=\frac{1}{\sqrt{1-K}} \, \Psi \, \sqrt{1-K}\\
&=-\frac{1}{1-K}\left(Q \sigma_L\right) \sigma_R
-\frac{1}{1-K}\left(Q \sigma_L\right)\frac{B}{1-K}\left(Q \sigma_R\right) \,,
\end{split}
\end{equation}
where $\sqrt{1-K}$ should be understood as a superposition of
wedge-based states:
\begin{equation}
\sqrt{1-K}=\frac{1-K}{\sqrt{1-K}}= \int_0^\infty dt \,
\frac{e^{-t}}{\sqrt{\pi t}} \, e^{t K} (1-K) \,.
\end{equation}
Since any function of $K$ is annihilated by the BRST operator,
$\Psi$ satisfies $Q \Psi +\Psi^2 = 0$
when $\widetilde{\Psi}$ satisfies $Q \widetilde{\Psi} +\widetilde{\Psi}^2 = 0$.
Although $\widetilde{\Psi}$ is simpler than $\Psi$,
the solution $\widetilde{\Psi}$ does not satisfy the reality condition
on the string field~\cite{Gaberdiel:1997ia}.
Let us demonstrate that $Q \widetilde{\Psi} +\widetilde{\Psi}^2 = 0$
follows only from~(\ref{KBsigma_1}) and~(\ref{KBsigma_2}).

\medskip
In calculating $Q \widetilde{\Psi}$,
we write $\widetilde{\Psi}$ as follows:
\begin{equation}
\label{Psi_Q}
\widetilde{\Psi}=-\frac{1}{1-K}\left(Q\sigma_L\right)\frac{1}{1-K} \, \sigma_R
-Q\left[\frac{1}{1-K}\left(Q \sigma_L\right)\frac{B}{1-K} \, \sigma_R \right] \,,
\end{equation}
where we used
\begin{equation}
Q \, \biggl[ \, \frac{B}{1-K} \, \biggr] = \frac{K}{1-K} = \frac{1}{1-K}-1 \,.
\end{equation}
The second term in (\ref{Psi_Q}) is BRST-exact and does not contribute to $Q \widetilde{\Psi}$.
The BRST transformation of $\widetilde{\Psi}$ is thus given by
\begin{equation}
\label{Q-Psi}
Q\widetilde{\Psi}=\frac{1}{1-K}(Q \sigma_L)\frac{1}{1-K} (Q \sigma_R) \,.
\end{equation}
Let us next calculate ${\widetilde{\Psi}}^2$. In this case we write $\widetilde{\Psi}$ as
\begin{equation}
\label{Psi-B-left}
\widetilde{\Psi} =\frac{1}{1-K}(Q \sigma_L)\frac{1}{1-K}(Q \sigma_R)B-\frac{1}{1-K}(Q \sigma_L)\frac{1}{1-K} \sigma_R\,(1-K)
\end{equation}
or as
\begin{equation}
\label{Psi-B-right}
\widetilde{\Psi} =\frac{B}{1-K}(Q \sigma_L)\frac{1}{1-K}(Q \sigma_R)+ \sigma_L\frac{1}{1-K} (Q \sigma_R) \, .
\end{equation}
Multiplying $\Psi$ in~(\ref{Psi-B-left}) and $\Psi$ in~(\ref{Psi-B-right}), we find
\begin{equation}
\label{Psi_B^2_bosonic}
\begin{split}
\widetilde{\Psi}^2 & =
\frac{1}{1-K}(Q \sigma_L)\frac{1}{1-K}(Q \sigma_R)B\sigma_L\frac{1}{1-K}(Q\sigma_R)\\
& \quad~
-\frac{1}{1-K}(Q \sigma_L)\frac{1}{1-K}\sigma_RB(Q \sigma_L)\frac{1}{1-K}(Q \sigma_R)\\
& \quad~
-\frac{1}{1-K}(Q \sigma_L)\frac{1}{1-K}\sigma_R(1-K) \sigma_L\frac{1}{1-K}(Q \sigma_R)\\
& = -\frac{1}{1-K}(Q \sigma_L)\frac{1}{1-K}(Q \sigma_R) \,,
\end{split}
\end{equation}
where we used
\begin{equation}
(Q \sigma_R)B\sigma_L-\sigma_RB(Q \sigma_L)=\,\{\,Q \sigma_R,B\,\}\, \sigma_L-B[(Q\sigma_R)\sigma_L+\sigma_R(Q\sigma_L)]=-[\,1-K,\sigma_R\,]\sigma_L \, .
\end{equation}
It follows from (\ref{Q-Psi}) and (\ref{Psi_B^2_bosonic})
that $Q\widetilde{\Psi}+\widetilde{\Psi}^2=0$ and therefore $\Psi$ in (\ref{bosonic_solutions_with_bcc}) satisfies the equation of motion~(\ref{bosonic-equation}).

\medskip
The solutions were originally constructed
by rewriting solutions for marginal deformations with regular operator products~\cite{Kiermaier:2010cf}.
While we do not necessarily assume that the modified boundary conditions are marginally connected
to the original ones,
the regularity conditions
imply
the existence of a marginal operator given by
$V(t)=\sigma_L\partial_t\sigma_R(t)$.\footnote{
We would like to thank the referee of this paper for explaining this to us.
}
We therefore need to relax the regularity conditions
for the construction of more general solutions.

\section{Solutions in open superstring field theory}
\label{section_super}
\setcounter{equation}{0}
In this section we construct solutions in open superstring field theory from bcc operators.
We first discuss the general prescription for the construction of solutions in open superstring field theory.
We then present solutions using bcc operators and discuss their regularity.
In the last subsection, we consider marginal deformations with regular operator products.

\subsection{Open superstring field theory}
The equation of motion for open superstring field theory in the NS sector~\cite{Berkovits:1995ab} is
\begin{equation}
\label{eom_for_super_SFT}
\eta_0\left(e^{-\Phi}\,Q\,e^\Phi\right)=0 \,,
\end{equation}
where $\Phi$ is the open superstring field.
It is Grassmann-even and has ghost number $0$ and picture number $0$.
The superconformal ghost sector is described by $\eta$, $\xi$, and $\phi$~\cite{Friedan:1985ge,Polchinski:1998rr},
and $\eta_0$ is the zero mode of $\eta$.
The BRST operator for the superstring is given by
\begin{equation}\label{Q_B}
Q= \int \Bigl[ \, \frac{dz}{2 \pi i} \, j_B (z)
- \frac{d \bar{z}}{2 \pi i} \, \tilde{\jmath}_B (\bar{z}) \, \Bigr]
\end{equation}
with
\begin{equation}
\begin{split}
j_B= c \, T_B^m + c \, T_B^{\eta \xi} + c \, T_B^\phi
+ \eta e^\phi \, T_F^m
+ b c \partial c
{}- b \eta \partial \eta e^{2 \phi} \,, \\
T_B^{\eta \xi} = {}- \eta \partial \xi \,, \qquad
T_B^\phi = {}-\frac{1}{2} \, \partial \phi \partial \phi
{}- \partial^2 \phi \,,
\end{split}
\end{equation}
where $T_B^m$ and $T_F^m$ are the holomorphic components
of the energy-momentum tensor
and the supercurrent, respectively, in the matter sector
and $\tilde{\jmath}_B$ is
the antiholomorphic counterpart of $j_B$.
Conformal normal ordering is implicit throughout this paper.
The superconformal ghost sector can also be described
by the free bosons $\chi$ and $\phi$
via the bosonization~\cite{Friedan:1985ge,Polchinski:1998rr}:
\begin{equation}
\xi(z) \cong e^{\chi}(z) \,, \quad \eta(z) \cong e^{-\chi}(z) \,.
\end{equation}
The operator product expansions are given by
\begin{equation}
\label{OPE_chi_eta}
\chi(z)\chi(0)\sim\ln z\,, \quad\phi(z)\phi(0)\sim-\ln z \, .
\end{equation}
We will use this description later.
The operator $\eta_0$ is 
a derivation with respect to the star product.
It is nilpotent and anticommutes with the BRST operator $Q$:
\begin{equation}
Q^2=0 \, ,\quad \eta_0^2=0 \, ,\quad \left\{Q,\eta_0\right\}=0\,.
\end{equation}
The gauge transformation of the string field is given by
\begin{equation}
e^\Phi\mapsto e^{\Phi^\prime}=\Omega\,e^{\Phi}\,\Lambda\quad
{\rm with}\quad Q \, \Omega=0 \, ,~~\eta_0\,\Lambda=0 \,.
\end{equation}

\subsection{General prescription}
\label{subsection_general_prescription}
For any solution $\Phi$ in open superstring field theory,
we can define a string field $\Psi$
of ghost number 1 and picture number 0 by
\begin{equation}
\label{Psi_Phi}
\Psi=e^{-\Phi}\,Q\,e^\Phi \, ,
\end{equation}
which satisfies the following equations:\footnote{
The first equation in~(\ref{super_cubic_eom}) 
takes the same form as the equation of motion
in the modified cubic
open superstring field theory~\cite{Preitschopf:1989fc, Arefeva:1989cp}.}
\begin{equation}
\label{super_cubic_eom}
Q\Psi+\Psi^2=0 \, ,\quad\eta_0\,\Psi=0 \, .
\end{equation}
Based on this fact,
Erler divided the construction of 
a solution in open superstring field theory into the following two steps~\cite{Erler:2007rh}:
the first step is to construct a string field $\Psi$ satisfying (\ref{super_cubic_eom}),
and the second step is to solve the equation (\ref{Psi_Phi}) for $\Phi$.
Note that different solutions for a given $\Psi$ are all gauge-equivalent.
For two solutions $\Phi_1$ and $\Phi_2$
we can show that $\Omega$ given by
\begin{equation}
e^{\Phi_1}=\Omega \, e^{\Phi_2}
\end{equation}
is annihilated by $Q$:
\begin{equation}
Q\, \Omega=Q(e^{\Phi_1} e^{-\Phi_2})=(e^{\Phi_1} \, \Psi) \, e^{-\Phi_2}
+e^{\Phi_1} \,( -\Psi \, e^{-\Phi_2})=0 \,,
\end{equation}
where we used
$Q \, e^{\Phi_1} = e^{\Phi_1} \Psi$ following from $e^{-\Phi_1} \, Q \, e^{\Phi_1} = \Psi$
and
$Q \, e^{-\Phi_2} = -\Psi \, e^{-\Phi_2}$ following from
$e^{-\Phi_2} \, Q \, e^{\Phi_2} = -Q( e^{-\Phi_2}) \, e^{\Phi_2} = \Psi$.

\subsubsection{The free theory}
Let us consider the second step of the strategy by Erler for the free theory.
Namely, for a string field $\Psi$ satisfying
\begin{equation}
Q \Psi =0 \, ,\quad\eta_0\,\Psi=0 \,,
\end{equation}
we would like to solve
\begin{equation}
\label{free-Phi}
\Psi= Q \, \Phi
\end{equation}
for $\Phi$.
This can be solved when $\Psi$ is in the Fock space
because the cohomology of $Q$ is trivial in the large Hilbert space.
This can be seen by
the existence of an operator $R(t)$ satisfying
\begin{equation}
Q\cdot R(t)=1 \,,
\end{equation}
and we choose
\begin{equation}
R(t)=-c\xi\partial\xi e^{-2\phi}(t) 
\end{equation}
in this paper.
Consider the operator corresponding to the BRST-closed state $\Psi$
in the state-operator mapping and denote it by $\Psi (\xi)$.
To solve~(\ref{free-Phi}), let us insert $R(t)$ to the left of $\Psi (0)$:
\begin{equation}
R(-\epsilon) \, \Psi (0)
\end{equation}
with $0 < \epsilon < 1$.
The BRST transformation of this pair of operators is $\Psi (0)$
because $Q \cdot R(-\epsilon) = 1$ and $Q \cdot \Psi (0) = 0$. 
Therefore, the state $\widehat{\Psi}_L$ corresponding to
$R(-\epsilon) \, \Psi (0)$
satisfies
\begin{equation}
Q \, \widehat{\Psi}_L = \Psi \,.
\end{equation}
We thus have a solution $\Phi = \widehat{\Psi}_L$
to~(\ref{free-Phi}).
Similarly, the BRST transformation of
\begin{equation}
 \Psi (0) \, R(\epsilon)
\end{equation}
with $0 < \epsilon < 1$
is $-\Psi (0)$,
and the state $\widehat{\Psi}_R$ corresponding to
$\Psi (0) \, R(\epsilon)$
satisfies
\begin{equation}
Q \, \widehat{\Psi}_R = -\Psi \,.
\end{equation}
We have another solution $\Phi = -\widehat{\Psi}_R$
to~(\ref{free-Phi}).

\medskip
The solutions $\widehat{\Psi}_L$ and $-\widehat{\Psi}_R$ depend
on the parameter $\epsilon$, but they are all gauge-equivalent.
For example, $\widehat{\Psi}_L$
with different values of $\epsilon$ are gauge-equivalent
because
\begin{equation}
R(-\epsilon_1) \, \Psi (0)
- R(-\epsilon_2) \, \Psi (0)
= Q \cdot [ \, R(-\epsilon_2) \, R(-\epsilon_1) \, \Psi (0) \, ] \,.
\end{equation}
We can similarly show the equivalence of $-\widehat{\Psi}_R$
with different values of $\epsilon$
and the equivalence between $\widehat{\Psi}_L$ and $-\widehat{\Psi}_R$.
When the limit $\epsilon \to 0$ of
$\widehat{\Psi}_L$ or $-\widehat{\Psi}_R$ is finite,
the resulting state is a solution in the Fock space,
while $\widehat{\Psi}_L$ and $-\widehat{\Psi}_R$ with $\epsilon > 0$ are not.

\subsubsection{The interacting theory}
\paragraph{-- Non-real solutions}
This construction can be generalized to the interacting theory
and to the state $\Psi$ outside the Fock space.
Suppose that the state $\Psi$ satisfying~(\ref{super_cubic_eom})
is made of wedge-based states
and we can insert $R(t)$ to the left of all the other operator insertions.
The resulting state $\widehat{\Psi}_L$ satisfies
\begin{equation}
\label{Psi-hat_L}
Q \, \widehat{\Psi}_L = (1+\widehat{\Psi}_L) \, \Psi
\end{equation}
because when the BRST operator acts on $R(t)$, we have $\Psi$,
and when the BRST operator acts on the other operator insertions,
we have $-Q \Psi = \Psi^2$ with the insertion of $R(t)$,
which is $\widehat{\Psi}_L \Psi$.
Similarly, suppose we can insert $R(t)$ to the right of all the other operator insertions.
The resulting state $\widehat{\Psi}_R$ satisfies
\begin{equation}
\label{Psi-hat_R}
Q \, \widehat{\Psi}_R = -\Psi \, (1+\widehat{\Psi}_R) \,.
\end{equation}
From these equations we observe that
\begin{equation}
\begin{split}
(1+\widehat{\Psi}_L)^{-1}Q(1+\widehat{\Psi}_L)&=(1+\widehat{\Psi}_L)^{-1}(1+\widehat{\Psi}_L)\,\Psi=\Psi\,,\\
(1+\widehat{\Psi}_R)\,Q(1+\widehat{\Psi}_R)^{-1}&
=-[ \, Q(1+\widehat{\Psi}_R) \,] \, (1+\widehat{\Psi}_R)^{-1}
=\Psi\,(1+\widehat{\Psi}_R)(1+\widehat{\Psi}_R)^{-1}=\Psi\,,
\end{split}
\end{equation}
and we can construct two solutions $\Phi_L$ and $\Phi_R$ to~(\ref{Psi_Phi}):
\begin{equation}
e^{\Phi_L} = 1+\widehat{\Psi}_L \,, \quad e^{-\Phi_R} = 1+\widehat{\Psi}_R \,,
\end{equation}
or
\begin{equation}
\Phi_L = \ln \, (1+\widehat{\Psi}_L) \,, \quad
\Phi_R = -\ln \, (1+\widehat{\Psi}_R) \,,
\end{equation}
where $\ln \, ( 1 + X )$ for a string field $X$ is defined by
\begin{equation}
\label{ln(1+X)}
\ln \, (1+X) \equiv \sum_{n=1}^\infty \frac{(-1)^{n-1}}{n} \, X^n \,.
\end{equation}
The solutions $\Phi_L$ and $\Phi_R$ depend
on the insertion point of $R(t)$ in $\widehat{\Psi}_L$ and $\widehat{\Psi}_R$,
and we can also use $R(z)$ and $\widetilde{R} (\bar{z})$ in the bulk
instead of $R(t)$ on the boundary.
However, all these solutions are gauge-equivalent
because by construction they give the same $\Psi$.

\paragraph{-- Real solutions}
These solutions do not generally satisfy the reality condition~\cite{Gaberdiel:1997ia}.
For the open superstring field, the condition is stated as $\Phi^\ddagger=-\Phi$,
where $X^\ddagger$ is the conjugate of $X$
defined by the combination of the Hermitian conjugation (hc)
and the inverse BPZ conjugation ($\text{bpz}^{-1}$):
\begin{equation}
X^\ddagger \equiv \text{bpz}^{-1} \circ \text{hc} \, (X) \,.
\end{equation}
The conjugation satisfies
\begin{eqnarray}
(Q X)^\ddagger &=& {}- (-1)^{|X|} \, Q X^\ddagger \,,
\label{conjugation-with-Q_B} \\
(X \, Y)^\ddagger &=& Y^\ddagger \, X^\ddagger \,,
\end{eqnarray}
where $|X|$ 
denotes the Grassmann property of the string field $X$:
it is $0$ mod $2$ for a Grassmann-even state
and $1$ mod $2$ for a Grassmann-odd state.
See~\cite{Kiermaier:2007ki} for a detailed discussion
on the reality condition
relevant to our case, in particular, about the insertion of $R(t)$.

\medskip
When $\Psi$ satisfies the condition $\Psi^\ddagger=\Psi$,
which is the reality condition for the string field
in bosonic string field theory,
we can construct $\widehat{\Psi}_L$ and $\widehat{\Psi}_R$
such that they are conjugate to each other: $\widehat{\Psi}_L^\ddagger=\widehat{\Psi}_R$.
This implies that 
\begin{equation}
\label{Lddagger=-R}
\left(e^{\Phi_L}\right)^\ddagger=e^{-\Phi_R}\quad{\rm or}\quad \Phi_L^\ddagger=-\Phi_R\,.
\end{equation}
Since $\Phi_L$ and $\Phi_R$ are solutions to (\ref{Psi_Phi}) for the same $\Psi$,
they are gauge-equivalent:
\begin{equation}
e^{\Phi_R}=\Omega\, e^{\Phi_L}\quad{\rm with}\quad Q\,\Omega=0 \,.  
\end{equation}
We also
observe that the gauge transformation parameter $\Omega$ is real:
\begin{equation}
\label{Omega-real}
\Omega^\ddagger=(e^{\Phi_R}e^{-\Phi_L})^\ddagger=e^{\Phi_R}e^{-\Phi_L}=\Omega\,.
\end{equation}
Using these properties, 
we can construct a solution $\Phi_{\rm real}$
satisfying the reality condition
from $\Phi_L$ and $\Phi_R$
by the following gauge transformations~\cite{Erler:2007rh}:
\begin{equation}
e^{\Phi_{\rm real}}=\Omega^{1/2}e^{\Phi_L}=\Omega^{-1/2}e^{\Phi_R} \, .
\end{equation}
Since $\Phi_{\rm real}$ is gauge-equivalent to the solutions $\Phi_L$ and $\Phi_R$,
it satisfies the equation of motion,
and its reality follows from (\ref{Lddagger=-R}) and (\ref{Omega-real}):
\begin{equation}
\left(e^{\Phi_{\rm real}}\right)^\ddagger e^{\Phi_{\rm real}}=(\Omega^{1/2}e^{\Phi_L})^\ddagger \,\Omega^{1/2}e^{\Phi_L}
=e^{-\Phi_R}\Omega \,e^{\Phi_L}=1\,.
\end{equation}
The precise definition of $\Phi_{\rm real}$
in terms of $\widehat{\Psi}_L$ and $\widehat{\Psi}_R$ is
\begin{equation}
e^{\Phi_{\rm real}}=\Omega^{1/2}e^{\Phi_L}=\frac{1}{\sqrt{e^{\Phi_L}e^{-\Phi_R}}}e^{\Phi_L}=\frac{1}{\sqrt{(1+\widehat{\Psi}_L)(1+\widehat{\Psi}_R)}}(1+\widehat{\Psi}_L) \, ,
\end{equation}
where
\begin{equation}
\frac{1}{\sqrt{1+X}} \equiv e^{-\frac{1}{2} \ln \, (1+X)}
\end{equation}
with $\ln \, (1+X)$ defined in~(\ref{ln(1+X)}).
More explicitly, we have
\begin{equation}
\frac{1}{\sqrt{(1+\widehat{\Psi}_L)(1+\widehat{\Psi}_R)}}=\sum_{n=0}^\infty
\frac{\Gamma(1/2)}{\Gamma(n+1)\,\Gamma(1/2-n)}\left(\widehat{\Psi}_L+\widehat{\Psi}_R+\widehat{\Psi}_L\widehat{\Psi}_R\right)^n \,.
\end{equation}
We can verify that
$Q \, ( \, \widehat{\Psi}_L+\widehat{\Psi}_R+\widehat{\Psi}_L\widehat{\Psi}_R \, ) = 0$
and $( \, \widehat{\Psi}_L+\widehat{\Psi}_R+\widehat{\Psi}_L\widehat{\Psi}_R \, )^\ddagger
= \widehat{\Psi}_L+\widehat{\Psi}_R+\widehat{\Psi}_L\widehat{\Psi}_R$,
consistent with the general discussion.
The real solution $\Phi_{\rm real}$ is expanded as follows:
\begin{equation}
\label{real_Phi_expanded_in_Psi}
\begin{split}
\Phi_{\rm real}=& \, \frac{1}{2}\left(\widehat{\Psi}_L-\widehat{\Psi}_R\right)-\frac{1}{4}\left(\widehat{\Psi}_L^2-\widehat{\Psi}_R^2\right)+\frac{1}{6}\left(\widehat{\Psi}_L^3-\widehat{\Psi}_R^3\right)-\frac{1}{24}\left(\widehat{\Psi}_L\widehat{\Psi}_R\widehat{\Psi}_L-\widehat{\Psi}_R\widehat{\Psi}_L\widehat{\Psi}_R\right)\\
&+\frac{1}{48}\left(\widehat{\Psi}_L^2\widehat{\Psi}_R-\widehat{\Psi}_L\widehat{\Psi}_R^2+\widehat{\Psi}_R\widehat{\Psi}_L^2-\widehat{\Psi}_R^2\widehat{\Psi}_L\right)+\ldots \,.
\end{split}
\end{equation}
We can confirm that $\Phi_{\rm real}$ satisfies the reality condition $\Phi_{\rm real}^\ddagger=-\Phi_{\rm real}$ 
for the first few terms in the expansion.

\bigskip
Each of the solutions $\Phi_L$, $\Phi_R$, and $\Phi_{\rm real}$ is defined
by a formal infinite sum,
and its convergence will depend on details of the solutions.
While this issue of convergence is specific to our particular constructions,
existence of $\Phi$ for a given $\Psi$ in general is an interesting question.
As we noted earlier, $\Psi$ satisfying (\ref{super_cubic_eom}) is a solution to the modified
cubic string field theory so that
this question is related to existence of a solution in the Berkovits
theory for a given solution in the modified cubic theory.
See~\cite{Erler:2010pr}
for recent discussion.

\bigskip
Finally, let us consider the action
for solutions constructed using this general prescription.
The action of open superstring field theory in the Berkovits formulation is given by~\cite{Berkovits:1995ab}
\begin{equation}
\label{action_Berkovits}
\begin{split}
&S=-\frac{1}{2g^2}\int_0^1dt \left[\partial_t\,{\rm Tr}\left(A_{\eta_0}A_Q\right)+{\rm Tr }\left(A_t\left\{A_Q,A_{\eta_0}\right\}\right)\right]\\
&{\rm with}\quad A_{\eta_0}=e^{-\Phi(t)}\,\eta_0e^{\Phi(t)}\, ,~~
A_{Q}=e^{-\Phi(t)}\,Qe^{\Phi(t)}\,,~~
A_{t}=e^{-\Phi(t)}\,\partial_te^{\Phi(t)}\, ,~~
\Phi(0)=0\,,~~
\Phi(1)=\Phi\,,
\end{split}
\end{equation}
where $g$ is the open string coupling constant and ${\rm Tr}(AB)=\langle A,B\rangle$.
The value of the action is gauge invariant
and the expression for the non-real solution $\Phi_L$
in terms of $\Psi$ and $\widehat{\Psi}_L$ is
\begin{equation}
\label{reduced_action}
S=\frac{1}{g^2}\sum_{n=3}^\infty\sum_{m=1}^{n-2}(-1)^{n-1}\frac{m}{n(n-1)}{\rm Tr}\left(\widehat{\Psi}_L^{m}\,(\eta_0\widehat{\Psi}_L)\,\widehat{\Psi}_L^{n-m-2}\,\Psi\right)\,.
\end{equation}
The derivation is given in appendix~\ref{appendix_action}.

\subsection{Construction of solutions from bcc operators}
\label{subsection-construction-from-bcc}
Our goal is to construct
solutions
to the equation of motion~(\ref{eom_for_super_SFT})
written in terms of bcc operators.
We can construct such solutions
if $\Psi$ in~(\ref{Psi_Phi}) is written in terms of bcc operators
and if we can insert $R(t)$ appropriately
to obtain $\widehat{\Psi}_L$ and $\widehat{\Psi}_R$
satisfying~(\ref{Psi-hat_L}) and~(\ref{Psi-hat_R}), respectively.
Since the $KB \sigma$ algebra also holds in open superstring field theory,\footnote{
As we mentioned in the introduction,
it is important that the $c$ ghost does not appear explicitly
because the BRST transformation of the $c$ ghost in the superstring
is different from that in the bosonic string.}
the solution (\ref{bosonic_solutions_with_bcc}) to the bosonic string equation of motion
\begin{equation}
\label{cubic_solutions_with_bcc}
\Psi=-\frac{1}{\sqrt{1-K}}\left(Q\sigma_L\right) \sigma_R\frac{1}{\sqrt{1-K}}-\frac{1}{\sqrt{1-K}}\left(Q \sigma_L\right)\frac{B}{1-K}\left(Q \sigma_R\right)\frac{1}{\sqrt{1-K}}
\end{equation} 
can be seen as a string field in superstring field theory satisfying $Q\Psi+\Psi^2=0$.
Furthermore,
$\Psi$ is annihilated by $\eta_0$
because all elements of
the $KB\sigma$ algebra
are in the small Hilbert space.
Therefore, $\Psi$ satisfies the two equations in (\ref{super_cubic_eom}).\footnote{
The string field $\Psi$ can be regarded
as a solution using bcc operators in the modified cubic theory.}
It also satisfies the reality condition $\Psi^\ddagger=\Psi$.

\medskip
Following the general prescription discussed in subsection~\ref{subsection_general_prescription}, we construct solutions using $\Psi$ in (\ref{cubic_solutions_with_bcc}) and the operator $R(t)$. We define $R$ by a state based on 
the wedge state $W_0$ of zero width with a local insertion of $R(t)$ on the boundary:
\begin{equation}
\langle \, \varphi, R \, \rangle
=\left\langle f\circ \varphi(0) \, R\left(\tfrac{1}{2}\right) \, \right\rangle_{C_{1}} \,.
\end{equation}
Using the state $R$,
we can construct $\widehat{\Psi}_L$ and $\widehat{\Psi}_R$ as follows:
\begin{equation}
\label{construction_of_hat_Psi}
\begin{split}
\widehat{\Psi}_L&=-\frac{1}{\sqrt{1-K}}R\left(Q \sigma_L\right) \sigma_R\frac{1}{\sqrt{1-K}}-\frac{1}{\sqrt{1-K}}R\left(Q \sigma_L\right)\frac{B}{1-K}\left(Q \sigma_R\right)\frac{1}{\sqrt{1-K}} \,, \\
\widehat{\Psi}_R&=\frac{1}{\sqrt{1-K}}\sigma_L\left(Q \sigma_R\right)R \frac{1}{\sqrt{1-K}}-\frac{1}{\sqrt{1-K}}\left(Q \sigma_L\right)\frac{B}{1-K}\left(Q \sigma_R\right)R\frac{1}{\sqrt{1-K}} \,.
\end{split}
\end{equation}
We therefore obtain
two non-real solutions $\Phi_L$ and $\Phi_R$ given by
\begin{equation}
\label{non_real_sbcc_solutions}
\begin{split}
e^{\Phi_L}&=1-\frac{1}{\sqrt{1-K}}R\left(Q \sigma_L\right) \sigma_R\frac{1}{\sqrt{1-K}}-\frac{1}{\sqrt{1-K}}R\left(Q \sigma_L\right)\frac{B}{1-K}\left(Q \sigma_R\right)\frac{1}{\sqrt{1-K}} \,, \\
e^{-\Phi_R}&=1+\frac{1}{\sqrt{1-K}}\sigma_L\left(Q \sigma_R\right)R \frac{1}{\sqrt{1-K}}-\frac{1}{\sqrt{1-K}}\left(Q \sigma_L\right)\frac{B}{1-K}\left(Q \sigma_R\right)R\frac{1}{\sqrt{1-K}}\,,
\end{split}
\end{equation}
and a real solution $\Phi_{\rm real}$ given by
\begin{equation}
\label{real_solution}
e^{\Phi_{\rm real}}=\frac{1}{\sqrt{(1+\widehat{\Psi}_L)(1+\widehat{\Psi}_R)}}(1+\widehat{\Psi}_L) \, .
\end{equation} 

\subsection{Regularity of the solutions}
\label{subsection_regularity}
In our construction of the solutions,
the operator $R(t)$ is inserted at the same point
where the BRST transformation of the bcc operator $\sigma_L$ or $\sigma_R$
is inserted.
Furthermore, in the expressions of $1/(1-K)$ and $1/\sqrt{1-K}$ in~(\ref{laplace-transf}),
the integral region of $t$ reaches $t=0$,
which corresponds to $W_0$ of zero width.
Therefore, various operator insertions in the solutions collide
and could make the solutions singular.
In this subsection, we investigate the regularity of the solutions
and demonstrate that
they do not have any short-distance singularity of operator products.

\medskip
Let us consider operator insertions in the superconformal ghost sector.
There are two sources of such operator insertions:
one is $R(t)=-c\xi\partial\xi e^{-2\phi}(t)$
and the other is the BRST operator
acting on the bcc operators
$\sigma_L (t)$ and $\sigma_R (t)$.
Since bcc operators behave as superconformal primary fields
under superconformal transformations,
their BRST transformations are determined by their weights.
{}From the regularity condition
\begin{equation}
\lim_{\epsilon \to +0} \sigma_L(t) \, \sigma_R(t+\epsilon)=1 \,,
\end{equation}
we assume that $\sigma_L (t)$ and $\sigma_R (t)$ behave
as superconformal primary fields of weight $0$.
Then their BRST transformations are given by
\begin{equation} 
\begin{split}
Q\cdot \sigma_{L}(t)
&=c\partial_t\sigma_{L}(t)+\eta e^{\phi}G_{-1/2}\cdot\sigma_{L}(t)\,, \\
Q\cdot \sigma_{R}(t)
&=c\partial_t\sigma_{R}(t)+\eta e^{\phi}G_{-1/2}\cdot\sigma_{R}(t)\,,
\end{split}
\end{equation}
where $G_{-1/2}$ generates the supersymmetry transformation:
\begin{equation}
G_{-1/2} \cdot \varphi(t) \equiv
\int_{C(t)} \Bigl[ \, \frac{dz}{2\pi i} \, T_F(z) \,
- \frac{d \bar{z}}{2 \pi i} \, \widetilde{T}_F(\bar{z}) \,
\Bigr] \, \varphi(t) \,.
\end{equation}
Here $T_F(z)$ and $\widetilde{T}_F(\bar{z})$ are
the holomorphic and antiholomorphic components, respectively,
of the world-sheet supercurrent,
and $C(t)$ is a contour
in the upper half-plane
which runs from the point $t+\epsilon$
on the real axis to the point $t-\epsilon$ on the real axis
in the limit $\epsilon \to 0$ with $\epsilon > 0$.

\medskip
In the description
using $\phi$ and $\chi$,
the operator $R(t)$ is written as
\begin{equation}
R(t) = c\,e^{2\chi} e^{-2\phi}(t)\, ,
\end{equation}
and
the BRST transformations of bcc operators are
\begin{equation}  
\label{BRST_of_sigma}
\begin{split}
&Q\cdot \sigma_{L}(t)
=c\partial_t\sigma_{L}(t)+e^{-\chi} e^{\phi}G_{-1/2}\cdot\sigma_{L}(t) \,, \\
&Q\cdot \sigma_{R}(t)
= c\partial_t\sigma_{R}(t)+e^{-\chi} e^{\phi}G_{-1/2}\cdot\sigma_{R}(t) \,.
\end{split}
\end{equation}
From these expressions,
we notice that the superconformal ghost sector of (\ref{cubic_solutions_with_bcc}) and (\ref{construction_of_hat_Psi}) has a simple structure. Namely, the superconformal ghosts $\chi$ and $\phi$ always appear in the form $e^{m\chi}e^{-m\phi}$. 
Because of the difference in the sign of the two operator product expansions in~(\ref{OPE_chi_eta}),
the operator product of $e^{m\chi}e^{-m\phi}(t)$ and $e^{n\chi}e^{-n\phi}(t)$ is regular
for arbitrary $m$ and $n$:
\begin{equation}
\label{superghost_OPE}
\lim_{\epsilon\rightarrow0}
e^{m\chi}e^{-m\phi}(t) \, e^{n\chi}e^{-n\phi}(t+\epsilon)
=(-1)^{mn}e^{(m+n)\chi}e^{-(m+n)\phi}(t) \, .
\end{equation}
We therefore find that the superconformal ghost sector does not have 
any short-distance singularity of operator products.
As the matter and $bc$ ghost sectors do not suffer from any singularity,
we conclude that the operator products in (\ref{cubic_solutions_with_bcc}) and (\ref{construction_of_hat_Psi}) are regular.

\bigskip
In the same way,
we can show that star products made of $\widehat{\Psi}_L$ and $\widehat{\Psi}_R$
are also regular.
This guarantees the regularity of operator products
in the solutions 
$\Phi_L$, $\Phi_R$, and $\Phi_{\rm real}$,
although there still remains the issue of convergence in the infinite sum
mentioned in
subsection~\ref{subsection_general_prescription}.

\subsection{Marginal deformations with regular operator products}
\label{subsection_marginal}
We conclude this section
by discussing the relation between our solutions constructed from bcc operators
and 
previous solutions for marginal deformations
 with regular operator products~\cite{ Erler:2007rh,Okawa:2007ri, Okawa:2007it},
where the regularity conditions of bcc operators are satisfied.
The marginal operator $V_1$ in the superstring is
the supersymmetry transformation
of a superconformal primary field $\widehat{V}_{1/2}$
in the matter sector
of weight $1/2$:
\begin{equation}\label{ssV1}
V_1(t) = G_{-1/2} \cdot \widehat{V}_{1/2}(t)\,. 
\end{equation}
An integrated vertex operator
in the $0$ picture
is an integral of $V_1$ on the boundary:
\begin{equation}
\int_a^b dt\, V_1(t)
= \int_a^b dt \, G_{-1/2} \cdot \widehat{V}_{1/2}(t) \,.
\end{equation}
It is invariant under the BRST transformation
up to nonvanishing terms from the endpoints
of the integral region:
\begin{equation}\label{QVab}
\begin{split}
Q \cdot \int_a^b dt\, V_1(t)
&= \int_a^b dt \,\partial_t \,
[ \, cV_1(t) + \eta e^\phi \, \widehat{V}_{1/2}(t) \, ]\\
&= [ \, cV_1(b) + \eta e^\phi \, \widehat{V}_{1/2}(b) \, ]
- [ \, cV_1(a) + \eta e^\phi \, \widehat{V}_{1/2}(a) \, ] \,.
\end{split}
\end{equation}
The operator $cV_1(t) + \eta e^\phi \, \widehat{V}_{1/2}(t)$
is annihilated by the BRST operator,
so it can be written as a BRST transformation of an operator.
Since
\begin{equation}\label{unintegrated-vertex-operator}
\lim_{\epsilon \to 0} R(t-\epsilon) \,
[ \, cV_1(t) + \eta e^\phi \, \widehat{V}_{1/2}(t) \, ]
= c \xi e^{-\phi}\widehat{V}_{1/2}(t) \,,
\end{equation}
we have
\begin{equation}\label{O1=QO1hat}
cV_1(t) + \eta e^\phi \, \widehat{V}_{1/2}(t)
= Q \cdot \mathcal{V} (t)
\end{equation}
with
\begin{equation}
\mathcal{V} (t) = c \xi e^{-\phi}\widehat{V}_{1/2}(t) \,.
\end{equation}

\medskip
When operator products of $V_1$ are regular,\footnote{
In the following,
we also need regularity conditions involving $\widehat{V}_{1/2}$.
A set of conditions, which is sufficient
for the solutions to be well defined,
can be stated in the following way:
the operators $V_1(t_1)V_1(t_2)^n$ and $\widehat{V}_{1/2}(t_1)V_1(t_2)^n$ 
are finite in the limit $t_1\rightarrow t_2$ for any positive integer $n$ and the operator $\widehat{V}_{1/2}(t_1)\widehat{V}_{1/2}(t_2)V_1(t_2)^n$
vanishes in the limit $t_1\rightarrow t_2$ for any positive integer $n$.
} 
bcc operators are given by 
\begin{equation}
\label{marginal-bcc}
\sigma_L(a) \, \sigma_R(b)
=\exp\left[\lambda\int_a^bdt \, V_1(t)\right]
=\exp\left[\lambda\int_a^bdt \, G_{-1/2}\cdot\widehat{V}_{1/2}(t)\right] \, ,
\end{equation}
where $\lambda$ is a deformation parameter.
We can confirm that the conditions described in subsection~\ref{subsection_KBsigma} are satisfied.
The BRST transformation of (\ref{marginal-bcc}) is given by
\begin{equation}
\label{Q_on_e}
Q \cdot \exp\left[\lambda\int_a^bdt \, V_1(t)\right]
=\exp\left[\lambda\int_a^bdt V_1(t)\right]\lambda \left(Q\cdot \mathcal{V} \right)(b)-\lambda \left(Q\cdot \mathcal{V} \right)(a) \exp\left[\lambda\int_a^bdt V_1(t)\right] \,.
\end{equation}
We can see that the effect of the BRST transformation is localized
at the endpoints $a$ and $b$, which is consistent with the local property of bcc operators.
Furthermore, the BRST transformations of the two terms
on the right-hand side of~(\ref{Q_on_e}) are
\begin{equation}
\label{BRST_marginal_2}
\begin{split}
Q\cdot \left[\lambda \left(Q\cdot \mathcal{V} \right)(a) \exp\left[\lambda\int_a^bdt V_1(t)\right]\right]&=-\lambda \left(Q\cdot \mathcal{V} \right)(a) \exp\left[\lambda\int_a^bdt V_1(t)\right]\lambda \left(Q\cdot \mathcal{V} \right)(b) \,, \\
Q\cdot \left[ \exp\left[\lambda\int_a^bdt V_1(t)\right]\lambda \left(Q\cdot \mathcal{V} \right)(b)\right]&=-\lambda \left(Q\cdot \mathcal{V} \right)(a) \exp\left[\lambda\int_a^bdt V_1(t)\right]\lambda \left(Q\cdot \mathcal{V} \right)(b) \,,
\end{split}
\end{equation}
where we used
\begin{equation}
\lim_{t_1 \to t_2}Q\cdot \mathcal{V} (t_1) \,\, Q\cdot \mathcal{V} (t_2)=0\,.
\end{equation}
From (\ref{Q_on_e}) and (\ref{BRST_marginal_2}), we identify
the BRST transformations
of bcc operators as
\begin{equation}
\begin{split}
\sigma_L(a)\left(Q\cdot\sigma_R\right)(b)&=\exp\left[\lambda\int_a^bdt V_1(t)\right]\lambda \left(Q\cdot \mathcal{V} \right)(b) \,, \\
\left(Q\cdot\sigma_L\right)(a)\sigma_R(b)&=-\lambda \left(Q\cdot \mathcal{V} \right)(a) \exp\left[\lambda\int_a^bdt V_1(t)\right] \,, \\
\left(Q\cdot\sigma_L\right)(a)\left(Q\cdot\sigma_R\right)(b)&=-\lambda \left(Q\cdot \mathcal{V} \right)(a) \exp\left[\lambda\int_a^bdt V_1(t)\right]\lambda \left(Q\cdot \mathcal{V} \right)(b) \,.
\end{split}
\end{equation}
When the operators are inserted on wedge states,
these relations can be translated as follows: 
\begin{equation}
\label{relations_states}
\begin{split}
\sigma_L\,e^{\alpha K}\left(Q\sigma_R\right)&=e^{\alpha(K+\lambda V_1)}\lambda \left(Q \mathcal{V} \right) \,, \\
\left(Q\sigma_L\right)\,e^{\alpha K}\,\sigma_R&=-\lambda \left(Q\mathcal{V} \right)e^{\alpha({K+\lambda V_1})} \,, \\
\left(Q\sigma_L\right)\,e^{\alpha K}\left(Q\sigma_R\right)&=-\lambda \left(Q \mathcal{V} \right) e^{\alpha(K+\lambda V_1)}\lambda \left(Q \mathcal{V} \right)\,,
\end{split}
\end{equation}
where the states $V_1$ and $\mathcal{V}$ are defined by the states based on the wedge state $W_0$ of zero width with a local insertion of $V_1(t)$ and $\mathcal{V}(t)$, respectively, on the boundary.
Note that $e^{\alpha(K+\lambda V_1)}$ is a wedge state with the modified boundary conditions~\cite{Kiermaier:2010cf}. 

\medskip
We can now write the solutions (\ref{non_real_sbcc_solutions})
in terms of $K$, $B$, $V_1$, and $\mathcal{V}$ as follows:
\begin{equation}
\begin{split}
e^{\Phi_L}&=1+\frac{1}{\sqrt{1-K}}\lambda\mathcal{V}\frac{1}{\sqrt{1-K}}+\frac{1}{\sqrt{1-K}}\lambda\mathcal{V}\frac{B }{1-K-\lambda V_1}\lambda \left( Q\mathcal{V}\right)\frac{1}{\sqrt{1-K}}\,,\\
e^{-\Phi_R}&=1-\frac{1}{\sqrt{1-K}}\lambda \mathcal{V}\frac{1}{\sqrt{1-K}}-\frac{1}{\sqrt{1-K}}\lambda\left( Q\mathcal{V}\right) \frac{B}{1-K-\lambda V_1}\lambda \mathcal{V}\frac{1}{\sqrt{1-K}}\,.
\end{split}
\end{equation}
Here we used the relations
\begin{equation}
R\,(Q\mathcal{V})=\mathcal{V}\,,\quad(Q\mathcal{V})R=-\mathcal{V}\,,
\end{equation}
which follow from (\ref{unintegrated-vertex-operator}).
It is straightforward 
to express a real solution (\ref{real_solution}) in terms of $K$, $B$, $V_1$, and $\mathcal{V}$.
As in the bosonic theory~\cite{Kiermaier:2010cf},
our solutions correspond to the special choice $f(K)=1/\sqrt{1-K}$ in the class of solutions by Erler~\cite{Erler:2007rh}.

\section{Gauge-invariant observables}
\label{section_gauge-invariant}
\setcounter{equation}{0}
A classical solution in open string field theory is expected to
correspond to a BCFT. 
In particular,
the solutions constructed from bcc operators
should correspond to the BCFT associated with the bcc operators.
In this section,
we provide evidence
by calculating
the gauge-invariant observables
introduced in \cite{Hashimoto:2001sm, Gaiotto:2001ji}.

\medskip
The gauge-invariant observable $W(\Psi,\mathcal{O})$ is defined
for a solution $\Psi$ in open bosonic string field theory
and an on-shell closed string vertex operator $\mathcal{O}$ of weight $(0,0)$~\cite{Hashimoto:2001sm, Gaiotto:2001ji}.
In~\cite{Ellwood:2008jh},
Ellwood conjectured that
it is related to the closed string one-point function on the disk
with the modified boundary conditions associated with the solution.
The relation is given by
\begin{equation}
\label{relation_Ellwood}
W(\Psi,\mathcal{O})=A_*(\mathcal{O})-A_0(\mathcal{O}) \, ,
\end{equation}
where $A_0(\mathcal{O})$ and $A_*(\mathcal{O})$ 
are the closed string one-point functions on
the unit disk
\begin{equation}
\label{disk_amplitude}
A_0(\mathcal{O})=\frac{1}{2\pi i}\langle \mathcal{O}(0)c(1)\rangle_{{\rm disk}}\,,
\quad
A_*(\mathcal{O})=\frac{1}{2\pi i}\langle \mathcal{O}(0)c(1)\rangle_{{\rm disk},\,{\rm BCFT}_*} \, ,
\end{equation}
in the original BCFT for $A_0(\mathcal{O})$
and in the BCFT$_\ast$ associated with the solution $\Psi$ for $A_\ast(\mathcal{O})$,
as indicated by the subscript.

\medskip
Based on this conjecture by Ellwood,
it is expected that the gauge-invariant observables for the solutions constructed from bcc operators
reproduce the one-point functions in the BCFT associated with the bcc operators.
After recalling some basic properties
of the gauge-invariant observables, we evaluate $W(\Psi,\mathcal{O})$
for the solution (\ref{bosonic_solutions_with_bcc}) in the bosonic string constructed in~\cite{Kiermaier:2010cf},
and then we extend the calculation
to the solutions (\ref{non_real_sbcc_solutions}) and (\ref{real_solution}) in the superstring.

\subsection{Properties of $W(\varphi,\mathcal{O})$}
For a state $\varphi$ in the Fock space,
$W(\varphi,\mathcal{O})$ is defined 
by the following correlator on the upper half-plane:
\begin{equation}
W(\varphi,\mathcal{O})=\langle \mathcal{O}(i)f_I\circ \varphi(0)\rangle_{\rm UHP} \quad{\rm with}\quad f_I(\xi)=\frac{2\xi}{1-\xi^2}\,.
\end{equation}
The conformal transformation $f_I (\xi)$ is associated with the wedge state $W_0$ of zero width,
which is also called the identity state:
\begin{equation}
\langle \, \varphi, W_0 \, \rangle
= \langle \, f \circ \varphi (0) \, \rangle_{C_1}
= \langle \, f_I \circ \varphi (0) \, \rangle_{\rm UHP} \,.
\end{equation}
Therefore, $W(\varphi,\mathcal{O})$ can be thought of as a BPZ inner product
of $\varphi$ and the identity state $W_0$
with an insertion of $\mathcal{O}$ at the conical singularity,
and we can formally express $W(\varphi,\mathcal{O})$
in the sliver frame as
\begin{equation}
\label{formal-expression}
W(\varphi,\mathcal{O})=\langle \, \mathcal{O}(i \infty) \, f \circ \varphi(0) \, \rangle_{C_1} \,.
\end{equation}
This expression is formal because the point $i \infty$ where $\mathcal{O}$ is inserted
is outside the coordinate patch, and we need another patch for a more rigorous treatment.

\medskip
It follows from the definition that
$W(\varphi,\mathcal{O})$ vanishes for any BRST-exact state and has a cyclic property:
\begin{eqnarray}
\label{BRST_W}
W(Q\Lambda,\mathcal{O})&=&0 \, ,\\
\label{similarity_transf_W}
W(\varphi_1\,\varphi_2,\mathcal{O})&=&(-1)^{|\varphi_1|\,|\varphi_2|}W(\varphi_2\,\varphi_1,\mathcal{O}) \, .
\end{eqnarray}
It is easy to see the cyclic property (\ref{similarity_transf_W}) from the expression (\ref{formal-expression}).
The observable $W(\Psi,\mathcal{O})$ is thus
invariant under gauge transformations in the bosonic string:
\begin{equation}
W\left(\delta\Psi,\mathcal{O}\right)=W\left(Q\Lambda+[\Psi,\Lambda],\mathcal{O}\right)=0\,.
\end{equation}

\medskip
We can also represent $W(\varphi,\mathcal{O})$ as a correlator on the unit disk
by the conformal transformation $f_{\rm disk}(\xi)$ from the upper half-plane to the disk:
\begin{equation}
W(\varphi,\mathcal{O})=\langle \mathcal{O}(0)f_{\rm disk}\circ f_I\circ \varphi(0)\rangle_{\rm disk}\quad{\rm with}\quad f_{\rm disk}(\xi)=\frac{1+i\xi}{1-i\xi} \, ,
\end{equation}
where the explicit form of $f_{\rm disk}\left(f_I(\xi)\right)$ is
\begin{equation}
f_{\rm disk}\left(f_I(\xi)\right)=\left(\frac{1+i\xi}{1-i\xi}\right)^2 \, .
\end{equation}
Consider a state $\Sigma$ based on $W_s$ defined by
\begin{equation}
\langle\varphi,\Sigma\rangle
=\Bigl\langle \, f\circ\varphi(0)\prod_{i=1}^n O_i(z_i) \, \Bigr\rangle_{C_{s+1}}
\quad{\rm with}\quad \frac{1}{2}\leq \Re z_i\leq s+\frac{1}{2} \, ,
\end{equation}
where $O_i(z_i)$ are some local operators.
A formal expression for $W(\Sigma,\mathcal{O})$ in the sliver frame
can be written as
\begin{equation}
W(\Sigma,\mathcal{O})
=\Bigl\langle \, \mathcal{O} (i \infty) \, \prod_{i=1}^n O_i(z_i) \, \Bigr\rangle_{C_s} \,,
\end{equation}
where the surface $C_s$ is represented in the region
$\frac{1}{2}\leq \Re z_i\leq s+\frac{1}{2}$ of the upper half-plane.
A precise expression for $W(\Sigma,\mathcal{O})$ can be obtained
by a conformal transformation from the surface $C_s$ to the unit disk:
\begin{equation}
\label{W_Sigma}
W(\Sigma,\mathcal{O})
=\Bigl\langle \, \mathcal{O}(0)\prod_{i=1}^n h_s\circ O_i(z_i) \, \Bigr\rangle_{\rm disk} \, ,
\end{equation} 
where the map $h_s(z)$ from $C_s$ to the unit disk is given by
\begin{equation}
\label{hs}
h_s(z)
= f_{\rm disk} \Bigl( f^{-1} \bigl( \, \tfrac{2z}{s}\, \bigr) \Bigr)
= \exp\frac{2\pi i z}{s} \, .
\end{equation}
In the following,
we use the expression of the gauge-invariant observables on the unit disk.

\subsection{Gauge-invariant observables in the bosonic string}
We start with the evaluation of the gauge-invariant observable
for the solution~(\ref{bosonic_solutions_with_bcc}) in the bosonic string
\begin{equation}
W(\Psi,\mathcal{O})\quad{\rm with}\quad\Psi=-\frac{1}{\sqrt{1-K}}\left(Q \sigma_L\right) \sigma_R \, \frac{1}{\sqrt{1-K}}
-\frac{1}{\sqrt{1-K}}\left(Q \sigma_L\right)\frac{B}{1-K}\left(Q \sigma_R\right)\frac{1}{\sqrt{1-K}} \,,
\end{equation}
where $\mathcal{O}=c\tilde{c}O_m$ with $O_m$ being a matter conformal primary field of weight $(1,1)$.
Using the cyclic property (\ref{similarity_transf_W}),
$W(\Psi,\mathcal{O})$ reduces to the form 
\begin{equation}
W(\Psi,\mathcal{O})=W(\widetilde{\Psi},\mathcal{O})\quad{\rm with}\quad\widetilde{\Psi}=-\frac{1}{1-K}\left(Q\sigma_L\right)\frac{1}{1-K} \, \sigma_R
-Q\left[\frac{1}{1-K}\left(Q \sigma_L\right)\frac{B}{1-K} \, \sigma_R \right] \,,
\end{equation}
where $\widetilde{\Psi}$ is defined 
in~(\ref{tilde-Psi})
and we wrote its expression given in~(\ref{Psi_Q}).
Since 
BRST-exact terms do not contribute to the gauge-invariant observables,
$W(\widetilde{\Psi},\mathcal{O})$ further reduces to the following form:
\begin{equation}
W(\widetilde{\Psi},\mathcal{O})=W(\widetilde{\Psi}_\ast,\mathcal{O})\quad{\rm with}\quad\widetilde{\Psi}_\ast=-\frac{1}{1-K}(Q\sigma_L)\frac{1}{1-K}\sigma_R \, .
\end{equation}
When $\sigma_L(t)$ and $\sigma_R(t)$ are matter primary fields of weight $0$,
their BRST transformations are
\begin{equation}
Q\cdot \sigma_{L}(t)=c\partial_t\sigma_{L}(t)\,, \quad
Q\cdot \sigma_{R}(t)=c\partial_t\sigma_{R}(t)\,,
\end{equation}
and $\widetilde{\Psi}_\ast$ is written as
\begin{equation}
\widetilde{\Psi}_\ast=-\frac{1}{1-K}c[K,\sigma_L]\frac{1}{1-K}\sigma_R\,,
\end{equation}
where $c$ is a state based on the wedge state $W_0$ of zero width with a local insertion of $c(t)$ on the boundary. 
Note that an insertion of $\partial_t\sigma_L(t)$
is translated into the commutator $[K,\sigma_L]$.
We can further
rewrite $\widetilde{\Psi}_\ast$ as
\begin{eqnarray}
\nonumber\widetilde{\Psi}_\ast&=&\frac{1}{1-K}[1-K,c\sigma_L]\frac{1}{1-K}\sigma_R+\frac{1}{1-K}[K, c]\sigma_L\frac{1}{1-K}\sigma_R\\
\nonumber&=&c\sigma_L\frac{1}{1-K}\sigma_R-\frac{1}{1-K}c+\frac{1}{1-K}[K,c]\sigma_L\frac{1}{1-K}\sigma_R\\
\nonumber
&=&\int_0^\infty ds\,e^{-s}\,c\sigma_Le^{sK}\sigma_R-\int_0^\infty ds\,e^{-s}\,e^{sK}c+\int_0^\infty ds_1\int_0^\infty ds_2\, e^{-(s_1+s_2)}e^{s_1K}[K,c]\sigma_Le^{s_2K}\sigma_R \\
\nonumber&=&\int_0^\infty ds\,e^{-s}\,c\sigma_Le^{sK}\sigma_R-\int_0^\infty ds\,e^{-s}\,e^{sK}c+\int_0^\infty ds\int_0^1 d\tau\, se^{-s}\,e^{(1-\tau)sK}[K,c]\sigma_Le^{\tau sK}\sigma_R \, .\\
\end{eqnarray}
The commutator $[K,c\,]$ corresponds to an insertion of $\partial_t c(t)$ on the boundary,
and the operator $\partial_t c(t)$ is transformed
under the conformal map $h_s(z)$ in (\ref{hs}) as
\begin{equation}
h_s\circ \partial_t c(t)=\partial_t\left(\frac{s}{2\pi i}e^{-2\pi it/s}c(e^{2\pi it/s})\right)=-i\partial_\theta\left(e^{-i\theta}c(e^{i\theta})\right)\,,
\end{equation}
where
$\theta=\frac{2\pi t}{s}$.
Using the expression of the gauge-invariant observables on the unit disk,
we~find
\begin{equation}
\label{bosonic_W}
\begin{split}
W(\widetilde{\Psi}_\ast,\mathcal{O})=\int_0^\infty& dse^{-s}\frac{s}{2\pi i}\lim_{\epsilon\rightarrow +0}\left\langle\mathcal{O}(0) c\sigma_L(1)\sigma_R(e^{(2\pi-\epsilon)i})\right\rangle_{\rm disk}-\int_0^\infty dse^{-s}\frac{s}{2\pi i}\langle\mathcal{O}(0) c(1)\rangle_{\rm disk}\\
&-i\int_0^\infty ds\int_0^1 d\tau \,se^{-s}\partial_\theta\left[e^{-i\theta}\left\langle\mathcal{O}(0) c(e^{i\theta})\sigma_L(1)\sigma_R(e^{2\pi i\tau})\right\rangle_{\rm disk}\right]\Big|_{\theta=0} \, .
\end{split}
\end{equation}
We can factorize the matter and ghost sectors of the correlator in the last term as
\begin{equation}
e^{-i\theta}\left\langle\mathcal{O}(0) c(e^{i\theta})\sigma_L(1)\sigma_R(e^{2\pi i\tau})\right\rangle_{\rm disk}=e^{-i\theta}\left\langle c\tilde{c}(0) c(e^{i\theta})\right\rangle^{bc}_{\rm disk}\left\langle O_m(0)\sigma_L(1)\sigma_R(e^{2\pi i\tau})\right\rangle^{\rm matter}_{\rm disk} \,.
\end{equation}
From the rotation invariance of the disk amplitude, we find
\begin{equation}
\partial_\theta\left[e^{-i\theta}\left\langle c\tilde{c}(0) c(e^{i\theta})\right\rangle^{bc}_{\rm disk}\right]=0 \, .
\end{equation}
Therefore, the last term in (\ref{bosonic_W}) vanishes,
 and the final expression of the gauge-invariant observable is
 \begin{equation}
\label{W_bosonic_completed}
W(\Psi,\mathcal{O})=W(\widetilde{\Psi}_\ast,\mathcal{O})=\frac{1}{2\pi i}\lim_{\epsilon\rightarrow +0}\left\langle\mathcal{O}(0) c\sigma_L(1)\sigma_R(e^{(2\pi-\epsilon)i})\right\rangle_{\rm disk}-\frac{1}{2\pi i}\langle\mathcal{O}(0) c(1)\rangle_{\rm disk} \, .
\end{equation}
The second term on the right-hand side is $A_0(\mathcal{O})$.
The first term is $A_\ast(\mathcal{O})$ because of the insertions of bcc operators:
\begin{equation}
\label{super_tadpole}
\frac{1}{2\pi i}\lim_{\epsilon\rightarrow +0}\left\langle\mathcal{O}(0) c\sigma_L(1)\sigma_R(e^{(2\pi-\epsilon)i})\right\rangle_{\rm disk}=\frac{1}{2\pi i}\langle \mathcal{O}(0)c(1)\rangle_{{\rm disk},\,{\rm BCFT}_*}=A_*(\mathcal{O}) \, .
\end{equation}
We have
thus 
obtained the relation (\ref{relation_Ellwood}) conjectured by Ellwood.

\subsection{Gauge-invariant observables in the superstring}
Let us extend the calculation
to the superstring.
For a solution $\Phi$ in superstring field theory,
the gauge-invariant observable is defined by $W(e^{-\Phi} Qe^\Phi,\mathcal{O})$~\cite{Ellwood:2008jh},
where $\mathcal{O}$ is an on-shell closed string vertex operator of weight $(0,0)$.
In the NS-NS sector
we can take the operator $\mathcal{O}$ to be
\begin{equation}
\mathcal{O}=(\xi+\tilde{\xi})c\tilde{c}e^{-\phi}e^{-\tilde{\phi}}
\widehat{O}_m \, ,
\end{equation}
where $\widehat{O}_m$
is a matter superconformal primary field of weight $(\frac{1}{2},\frac{1}{2})$.
The conjecture by Ellwood in the superstring can be stated as~\cite{Ellwood:2008jh}
\begin{equation}
\label{Ellwood_super}
W(e^{-\Phi} Qe^\Phi,\mathcal{O})=A_*(\mathcal{O})-A_0(\mathcal{O}) \, .
\end{equation}
For the solutions 
(\ref{non_real_sbcc_solutions}) and (\ref{real_solution}),
the gauge-invariant observables are 
\begin{equation}
W(e^{-\Phi}Qe^\Phi,\mathcal{O})=W(\Psi,\mathcal{O}),
\end{equation}
where $\Psi$ is given by (\ref{cubic_solutions_with_bcc}).
As in the case of the bosonic string,
the calculation of the observables reduces to the following form: 
\begin{equation}
W(e^{-\Phi}Qe^\Phi,\mathcal{O})=W(\widetilde{\Psi}_\ast,\mathcal{O})\quad{\rm with}\quad\widetilde{\Psi}_\ast=-\frac{1}{1-K}(Q\sigma_L)\frac{1}{1-K}\sigma_R \, .
\end{equation}
The explicit form of $Q\cdot\sigma_L(t)$ in the superstring is given by
\begin{equation}
\label{tilde_Psi_1}
Q\cdot\sigma_L(t)=c\partial_t\sigma_L(t)+\eta e^\phi(G_{-1/2}\cdot\sigma_L)(t)
\end{equation}
when $\sigma_L(t)$ and $\sigma_R(t)$ are matter superconformal primary fields of weight $0$.
Since the $bc$ ghost sector of $\mathcal{O}$ is $c\tilde{c}$,
the second term
on the right-hand side of~(\ref{tilde_Psi_1})
does not contribute to the observables
because it does not contain any $c$ ghost.
In the case of the R-R sector,
the explicit form of $\mathcal{O}$ is not simple~\cite{Ellwood:2008jh}.
However, the $bc$ ghost sector is $c\tilde{c}$
so that the second term
on the right-hand side of~(\ref{tilde_Psi_1})
does not contribute to the observables either.
The calculations in both sectors
therefore reduce to that in the bosonic string,
and we exactly obtain the relation (\ref{Ellwood_super}) conjectured by Ellwood.

\section{Discussion}
\label{discussion}
\setcounter{equation}{0}
\paragraph{-- Universal coefficients}
In this paper,
we constructed a class of analytic solutions of open superstring field theory in the Berkovits formulation
from boundary condition changing operators satisfying the regularity conditions described in subsection~\ref{subsection_KBsigma}.
In the case of the solutions in the bosonic string,
the dependence on the matter sector is simple,
and we only need the information of three-point functions
with a pair of bcc operators and an arbitrary operator in the original BCFT~\cite{Kiermaier:2010cf}.
In the case of the solutions in the superstring,
the dependence on the matter sector is again simple.
All the solutions $\Phi_L$, $\Phi_R$, and $\Phi_{\rm real}$ are constructed from star products
of $\widehat{\Psi}_L$ and $\widehat{\Psi}_R$,
which are specified by giving $\langle \varphi, \widehat{\Psi}_L\rangle$ and $\langle \varphi, \widehat{\Psi}_R\rangle$
for an arbitrary state $\varphi$ in the Fock space, 
and the dependence of $\langle \varphi, \widehat{\Psi}_L\rangle$ and $\langle \varphi, \widehat{\Psi}_R\rangle$
on the matter sector reduces to the three-point functions
\begin{equation}
\label{matter-three-point}
\begin{split}
&\left\langle \varphi_m(t_1)(G_{-1/2}\cdot \sigma_L)(t_2)\sigma_R(t_3)\right\rangle^{\rm matter}_{\rm UHP}\,,\\
&\left\langle \varphi_m(t_1) \sigma_L(t_2)(G_{-1/2}\cdot\sigma_R)(t_3)\right\rangle^{\rm matter}_{\rm UHP}\,,\\
&\left\langle \varphi_m(t_1)(G_{-1/2}\cdot \sigma_L)(t_2)(G_{-1/2}\cdot\sigma_R)(t_3)\right\rangle^{\rm matter}_{\rm UHP}\,,
\end{split}
\end{equation}
where $\varphi_m(t)$ is an arbitrary matter conformal primary field.
For example, 
$\langle \varphi, \widehat{\Psi}_L\rangle$  for $\varphi(t)=-c\partial c e^{-\phi}\varphi_m(t)$ is given by 
\begin{equation} 
\label{varphi_Psihat_L}
\langle\varphi,\widehat{\Psi}_L\rangle=C_{\varphi}\, g(h)\quad {\rm with}\quad C_{\varphi}=\left\langle \varphi_m(0)(G_{-1/2}\cdot \sigma_L)(1)\sigma_R(\infty)\right\rangle^{\rm matter}_{\rm UHP}\,,
\end{equation}
and $g(h)$ is a universal function of the weight $h$ of $\varphi_m$,
which does not depend on the particular choice of $\varphi_m$ or the bcc operators $\sigma_L$ and $\sigma_R$.
The explicit form of $g(h)$ is
\begin{equation}
\begin{split}
g(h)=\left(h-\frac{1}{2}\right)\int_{\frac{1}{2}}^\infty dx\int_{0}^\infty ds\int_{\frac{1}{2}}^\infty &dy\,\frac{e^{1-x-s-y}}{2\pi \sqrt{(x-\frac{1}{2})(y-\frac{1}{2})}}\left(\frac{2\sin\theta_{s}}{L\sin\theta_x\sin\theta_{x+s}}\right)^{h+\frac{1}{2}} \\ 
&\times\frac{\theta_y\sin^2\theta_x+\theta_x\sin^2\theta_y-\sin\theta_x\sin\theta_s\sin\theta_y}{\sin^{2}\theta_s}
\end{split}
\end{equation}
for $h\neq\frac{1}{2}$ and $g(\frac{1}{2})=1$,
where $L=x+s+y$
and $\theta_\ell=\frac{\ell}{L}\pi$.
The details of the calculation are presented in appendix \ref{appendix_universal}.
It is straightforward to calculate
$\langle \varphi, \widehat{\Psi}_L\rangle$ and $\langle \varphi, \widehat{\Psi}_R\rangle$
for a different~$\varphi$.

\paragraph{-- Towards the generalization to singular bcc operators}
For a systematic construction
of solutions in open string field theory
for a given BCFT,
we would like to generalize the construction of solutions
to the case when operator products of the bcc operators are singular.\footnote{
See~\cite{Ellwood:2009zf,Bonora:2010hi,Erler:2011tc}
for other interesting approaches to this problem.
}
We expect that
the structure we observed
in the calculation of the gauge-invariant observables
in section~\ref{section_gauge-invariant}
gives us some insight into this generalization.
The solution takes the form
\begin{equation}
\widetilde{\Psi}=c\sigma_L\frac{1}{1-K}\sigma_R-\frac{1}{1-K}c+(\,{\rm irrelevant\,\, terms}\,) \,,
\end{equation}
where the irrelevant terms are the terms which do not contribute to the gauge-invariant observables
such as BRST-exact terms
or those containing 
$\partial_t c$.
From this structure,
it is obvious that the gauge-invariant observables
for the solution reproduce
the expected change of boundary conditions of the disk.

\medskip
We also observe this structure in the phantomless solution for tachyon condensation
by Erler and Schnabl~\cite{Erler:2009uj}.
It is given by
\begin{equation}
\label{phantomless}
\Psi=-\frac{1}{\sqrt{1-K}}\left(c-cKBc\right)\, \frac{1}{\sqrt{1-K}}\,,
\end{equation}
and the associated $\widetilde{\Psi}$ has the same structure:
\begin{equation}
\widetilde{\Psi}=\frac{1}{\sqrt{1-K}} \, \Psi \, \sqrt{1-K}=-\frac{1}{1-K}c+Q\left(\frac{1}{1-K}Bc\right)\,.
\end{equation}
In~\cite{Erler:2009uj} a class of gauge conditions called dressed $\mathcal{B}_0$ gauges
were introduced.
In fact the solutions (\ref{phantomless}) and (\ref{bosonic_solutions_with_bcc})
satisfy the same gauge condition in the class of dressed $\mathcal{B}_0$ gauges.
In the generalization to bcc operators with singular operator products,
the same structure may continue to hold,
and this gauge condition may play an important role.

\bigskip
\noindent
{\bf \large Acknowledgments}

\medskip
We would like to thank Michael Kiermaier, Michael Kroyter, and Pablo Soler for valuable comments
on an earlier version of the paper.
We also thank
the Yukawa Institute for Theoretical Physics at Kyoto University for  hospitality
during the workshop YITP-W-11-05 on ``Quantum Field Theory and String Theory'',
where T.N. presented the results in this paper.
We also benefited from the workshop ``Summer Institute 2011 Cosmology and String",
where this work was completed.
The work of T.N. was supported in part by JSPS Grant-in-Aid for JSPS Fellows.
The work of Y.O. was supported in part
by Grant-in-Aid for Young Scientists~(B) No.~21740161 from
the Ministry of Education, Culture, Sports, Science and Technology
(MEXT) of Japan,
by Grant-in-Aid for Scientific Research~(B) No.~20340048 from
the Japan Society for the Promotion of Science (JSPS),
and
by the Japan Society for the
Promotion of Science (JSPS) and the Academy of Sciences of the Czech
Republic (ASCR) under the Japan-Czech Republic Research Cooperative Program.

\appendix
\section{Evaluating the action}
\setcounter{equation}{0}
\label{appendix_action}
We discuss the structure of the action for 
solutions
constructed
by the general prescription explained
in subsection~\ref{subsection_general_prescription}.
As shown in~\cite{Berkovits:2004xh}, the action~(\ref{action_Berkovits})
of open superstring field theory in the Berkovits formulation can be written as
\begin{equation}
\label{simple_action}
S=-\frac{1}{g^2}\int_0^1dt\,{\rm Tr}\left(\left(\eta_0 A_t\right)A_Q\right)\,.
\end{equation}
Since the solutions $\Phi_L$, $\Phi_R$, and $\Phi_{\rm real}$ in subsection~\ref{subsection_general_prescription}
are gauge-equivalent,
the values of the action for them are the same.
Here we evaluate the action for $\Phi_L=\ln (1+\widehat{\Psi}_L)$.
Choosing the $t$-dependence of $\Phi(t)$ to be
\begin{equation}
e^{\Phi(t)}=1+t\widehat{\Psi}_L\,,
\end{equation}
we have
\begin{equation}
\label{A_t_A_Q}
\begin{split}
A_t&=(1+t\widehat{\Psi}_L)^{-1}\widehat{\Psi}_L\,,\\
A_Q&=(1+t\widehat{\Psi}_L)^{-1}(1+\widehat{\Psi}_L)\,t\Psi\,,
\end{split}\
\end{equation}
and
\begin{equation}
\label{eta_0_A_t}
\begin{split}
\eta_0 A_t&=-(1+t\widehat{\Psi}_L)^{-1}(t\,\eta_0 \widehat{\Psi}_L)(1+t\widehat{\Psi}_L)^{-1}\widehat{\Psi}_L+(1+t\widehat{\Psi}_L)^{-1}(\eta_0 \widehat{\Psi}_L)\\
&=(1+t\widehat{\Psi}_L)^{-1}(\eta_0 \widehat{\Psi}_L)\left[1+(1+t\widehat{\Psi}_L)^{-1}(-t\widehat{\Psi}_L)\right]\\
&=(1+t\widehat{\Psi}_L)^{-1}(\eta_0 \widehat{\Psi}_L)(1+t\widehat{\Psi}_L)^{-1}\,.
\end{split}
\end{equation}
Therefore, the action for the solution $\Phi=\Phi_L$ is
\begin{equation}
\label{action_for_g_L}
S=-\frac{1}{g^2}\int_0^1dt\,{\rm Tr}\left[(1+t\widehat{\Psi}_L)^{-1}(\eta_0\widehat{\Psi}_L)(1+t\widehat{\Psi}_L)^{-1}(1+t\widehat{\Psi}_L)^{-1}(1+\widehat{\Psi}_L)t\Psi\right]\,.
\end{equation}
The expansion of the action in $\Psi$ and $\widehat{\Psi}_L$ is given by
\begin{eqnarray}
\nonumber
S&=&-\frac{1}{g^2}\sum_{a,b=0}^{\infty}\int_0^1dt\,t^{a+b+1}\,{\rm Tr}\left[(-\widehat{\Psi}_L)^a(\eta_0\widehat{\Psi}_L)(b+1)(-\widehat{\Psi}_L)^{b}(1+\widehat{\Psi}_L)\Psi\right]\\
\label{action_in_Psi_1}
\nonumber&=&-\frac{1}{g^2}\sum_{a,b=0}^{\infty}\frac{b+1}{a+b+2}\,{\rm Tr}\left[(-\widehat{\Psi}_L)^a(\eta_0\widehat{\Psi}_L)(-\widehat{\Psi}_L)^{b}\Psi-(-\widehat{\Psi}_L)^a(\eta_0\widehat{\Psi}_L)(-\widehat{\Psi}_L)^{b+1}\Psi\right]\\
\nonumber&=&-\frac{1}{g^2}\sum_{a,b=0}^{\infty}\frac{a+1}{(a+b+2)(a+b+1)}\,{\rm Tr}\left[(-\widehat{\Psi}_L)^a(\eta_0\widehat{\Psi}_L)(-\widehat{\Psi}_L)^{b}\Psi\right]\\
\label{action_in_Psi_12}
&=&\frac{1}{g^2}\sum_{n=2}^\infty\sum_{m=0}^{n-2}(-1)^{n-1}\frac{m+1}{n(n-1)}\,{\rm Tr}\left[\widehat{\Psi}_L^{m}\,(\eta_0\widehat{\Psi}_L)\,\widehat{\Psi}_L^{n-m-2}\,\Psi\right]\,.
\end{eqnarray}
Since ${\rm Tr} \, [ \, \eta_0 ( A B ) \, ] = 0$, which follows from the BPZ property
$\langle \, \eta_0 A, B \, \rangle = -(-1)^{|A|} \langle \, A, \eta_0 B \, \rangle$,
we find that
\begin{equation}
\sum_{m=0}^{n-2}{\rm Tr}\left[\widehat{\Psi}_L^{m}\,(\eta_0\widehat{\Psi}_L)\,\widehat{\Psi}_L^{n-m-2}\,\Psi\right]={\rm Tr}\left[\eta_0\,(\widehat{\Psi}_L^{n-1}\,\Psi)\right]=0\,,
\end{equation}
where we also used $\eta_0 \Psi = 0$. 
We therefore obtain
\begin{equation}
\label{action_in_Psi_2}
S=\frac{1}{g^2}\sum_{n=3}^\infty\sum_{m=1}^{n-2}(-1)^{n-1}\frac{m}{n(n-1)}\,{\rm Tr}\left[\widehat{\Psi}_L^{m}\,(\eta_0\widehat{\Psi}_L)\,\widehat{\Psi}_L^{n-m-2}\,\Psi\right]\,.
\end{equation}
The first few terms in the expansion are
\begin{equation}
S=\frac{1}{g^2}\,{\rm Tr}\left[\frac{1}{6}\widehat{\Psi}_L(\eta_0\widehat{\Psi}_L)\Psi-\frac{1}{12}\left\{\widehat{\Psi}_L(\eta_0\widehat{\Psi}_L)\widehat{\Psi}_L\Psi+2\widehat{\Psi}_L^2(\eta_0\widehat{\Psi}_L)\Psi\right\}+\ldots\right]\,.
\end{equation}

\medskip
The state $\widehat{\Psi}_L$ is constructed by inserting $R(t)$ to $\Psi$,
and the operator $\eta_0$ annihilates $\Psi$
and changes $R(t)$ to the picture-lowering operator $Y(t)=c\partial \xi e^{-2\phi}(t)$.
Schematically,
each term in~(\ref{action_in_Psi_2}) consists of
$n$ states of $\Psi$, $n-2$ insertions of $R(t)$, and one $Y(t)$.

\section{Universal coefficients}
\setcounter{equation}{0}
\label{appendix_universal}
In this appendix
we evaluate
$\langle \varphi, \widehat{\Psi}_L\rangle$
for $\varphi(t)=-c\partial c e^{-\phi}\varphi_m(t)$,  
where $\varphi_m(t)$ is a matter conformal primary field of weight $h$.
When the bcc operators are superconformal primary fields of weight~$0$,
$\widehat{\Psi}_{L}$ is given by
\begin{equation}
\widehat{\Psi}_{L}=\widehat{\Psi}_{L1}+\widehat{\Psi}_{L2}+\widehat{\Psi}_{L3}
\end{equation}
with
\begin{equation}
\label{Psi_L123}
\begin{split}
\widehat{\Psi}_{L1}&=-\frac{1}{\sqrt{1-K}}ce^\chi e^{-\phi}\left(G_{-1/2} \sigma_L\right) \sigma_R\frac{1}{\sqrt{1-K}}\,,\\ 
\widehat{\Psi}_{L2}&=-\frac{1}{\sqrt{1-K}}ce^\chi e^{-\phi}\left(G_{-1/2} \sigma_L\right)\frac{B}{1-K}c\,[K, \sigma_R]\frac{1}{\sqrt{1-K}}\,,\\
\widehat{\Psi}_{L3}&=-\frac{1}{\sqrt{1-K}}ce^\chi e^{-\phi}\left(G_{-1/2} \sigma_L\right)\frac{B}{1-K}e^{-\chi} e^{\phi}\left(G_{-1/2} \sigma_R\right)\frac{1}{\sqrt{1-K}}\,, 
\end{split}
\end{equation}
where the states 
$e^{m\chi}$ and $e^{n\phi}$ are defined
by the states based on the wedge state $W_0$ of zero width
with a local insertion of $e^{m\chi}(t)$ and $e^{n\phi}(t)$, respectively, on the boundary.
For $\varphi(t)=-c\partial c e^{-\phi}\varphi_m(t)$, 
the inner product $\langle \varphi, \widehat{\Psi}_{L3} \rangle$ vanishes
because of the constraint from the $bc$ ghost number:
\begin{equation}
\langle \varphi, \widehat{\Psi}_L\rangle=\langle\varphi,\widehat{\Psi}_{L1}\rangle+\langle\varphi,\widehat{\Psi}_{L2}\rangle\,.
\end{equation}
We first evaluate the second term $\langle\varphi,\widehat{\Psi}_{L2}\rangle$
given by
\begin{equation}
\label{varphi_Psihat2}
\begin{split}
\langle\varphi,\widehat{\Psi}_{L2}\rangle=-\int_{\frac{1}{2}}^\infty dx\int_{0}^\infty ds\int_{\frac{1}{2}}^\infty dy&\frac{e^{1-x-s-y}}{\pi \sqrt{(x-\frac{1}{2})(y-\frac{1}{2})}}\\
&\times\langle f \circ\varphi(0)c e^\chi e^{-\phi}(G_{-1/2}\cdot \sigma_L)(x)\mathcal{B}
c\partial_s\sigma_R(x+s)
\rangle_{C_{L}}\,,
\end{split}
\end{equation}
where $L=x+s+y$ and the operator $\mathcal{B}$ is a line integral of the $b$ ghost
defined by
\begin{equation}
\label{mathcal_B_def}
\mathcal{B}=\int^{-i\infty}_{i\infty}\frac{dz}{2\pi i}b(z)\,.
\end{equation}
The correlator in (\ref{varphi_Psihat2}) can be factorized as follows:\footnote{
The normalization of the correlator on the upper half-plane is
$
\langle e^\chi \,c\partial c\partial^2 c\,e^{-2\phi}(z)\rangle_{\rm UHP}=-2
$ up to a factor of the spacetime volume.
We also use the following convention of factorization:
$\langle e^\chi\, c\partial c\partial^2 c\,e^{-2\phi}(z)\rangle_{\rm UHP}=-\langle c\partial c\partial^2 c(z)\rangle_{\rm UHP}^{bc}\langle e^\chi e^{-2\phi}(z)\rangle^{\chi\phi}_{\rm UHP}\langle 1\rangle^{\rm matter}_{\rm UHP}$
with $\langle c\partial c\partial^2 c(z)\rangle_{\rm UHP}^{bc}=-2$
and
$\langle e^\chi e^{-2\phi}(z)\rangle^{\chi\phi}_{\rm UHP}=-1$.
}
\begin{equation}
\begin{split}
\label{factorized_correlator_Phihat2}
\langle f \circ\varphi(0)c e^\chi e^{-\phi}(G_{-1/2}\cdot \sigma_L)(x)\mathcal{B}c
\partial_s\sigma_R
&(x+s)\rangle_{C_{L}}\\
=-\left(\frac{\pi}{2}\right)^{\frac{1}{2}}\langle c\partial c(0)c (x)\mathcal{B}c(x+s)&\rangle^{bc}_{C_{L}}\langle e^{-\phi}(0) e^\chi e^{-\phi}(x)\rangle^{\chi\phi}_{C_{L}}\\ 
\times\partial_b\langle f \circ\varphi_m(0)(G_{-1/2}\cdot &\sigma_L)(x)\sigma_R(b)\rangle^{\rm matter}_{C_L}\Bigr|_{b=x+s}\,. 
\end{split}
\end{equation}
Since $\varphi_m$, $G_{-1/2}\cdot \sigma_L$, and $\sigma_R$ are primary fields
of weight $h$, $\frac{1}{2}$, and $0$, respectively, we find
\begin{equation}
\begin{split}
\left\langle f \circ\varphi_m(0)\right.&(G_{-1/2}\cdot\left. \sigma_L)(a)\sigma_R(b)\right\rangle^{\rm matter}_{C_{L}}\\
&=\left(\frac{2}{L}\right)^h\left(\frac{\pi}{L\cos^2\theta_a}\right)^{\frac{1}{2}}\left\langle \varphi_m(0)(G_{-1/2}\cdot \sigma_L)(\tan\theta_a)\sigma_R(\tan \theta_b)\right\rangle^{\rm matter}_{\rm UHP}\\
&=C_{\varphi}\left(\frac{2}{L}\right)^h\left(\frac{\pi}{L}\right)^{\frac{1}{2}}\frac{1}{\sin\theta_a}\left(\frac{\sin\theta_{b-a}}{\sin\theta_a\sin\theta_b}\right)^{h-\frac{1}{2}}\,,
\end{split}
\end{equation}
where 
$0<a<b<L$, 
$\theta_\ell=\frac{\ell}{L}\pi$, 
and $C_{\varphi}$ is a constant independent of $a$, $b$, and $L$.
It is related to the coefficient of the matter three-point
function of $\varphi_m$, $G_{-1/2}\cdot \sigma_L$, and $\sigma_R$ as follows:
\begin{equation}
\left\langle \varphi_m(t_1)(G_{-1/2}\cdot \sigma_L)(t_2)\sigma_R(t_3)\right\rangle^{\rm matter}_{\rm UHP}=C_\varphi\frac{|t_3-t_2|^{h-\frac{1}{2}}}{|t_2-t_1|^{h+\frac{1}{2}}\,|t_3-t_1|^{h-\frac{1}{2}}}\,.
\end{equation}
In other words,
$C_{\varphi}$ is the matter three-point function with operators $\varphi_m$, $G_{-1/2}\cdot \sigma_L$, and $\sigma_R$ inserted at $0$, $1$,
and $\infty$, respectively:\footnote{
Since the weight of $\sigma_R$ vanishes,
we can simply send the position of $\sigma_R$ to infinity
without considering the conformal transformation $I(\xi)=-1/\xi$.
}
\begin{equation}
\label{matter-3-points}
C_{\varphi}=\left\langle \varphi_m(0)(G_{-1/2}\cdot \sigma_L)(1)\sigma_R(\infty)\right\rangle^{\rm matter}_{\rm UHP}\,.
\end{equation}
Therefore, the matter correlator in (\ref{factorized_correlator_Phihat2}) is
\begin{equation}
\begin{split}
\partial_b\langle f& \circ\varphi_m(0)(G_{-1/2}\cdot \sigma_L)(x)\sigma_R(b)\rangle^{\rm matter}_{C_L}\Bigr|_{b=x+s}\\
&=\left(h-\frac{1}{2}\right)C_{\varphi}\left(\frac{2}{L}\right)^h\left(\frac{\pi}{L}\right)^{\frac{3}{2}} 
\frac{1}{\sin\theta_{x+s}\sin\theta_{s}}\left(\frac{\sin\theta_{s}}{\sin\theta_x\sin\theta_{x+s}}\right)^{h-\frac{1}{2}}\,.
\end{split}
\end{equation}
The ghost sector correlators take the form
\begin{equation}
\begin{split}
\langle c\partial c(0)c (x)\mathcal{B}c(x+s)\rangle^{bc}_{C_L}&=\frac{L^2}{\pi^3}\left(-\theta_y\sin^2\theta_x-\theta_x\sin^2\theta_y+\sin\theta_x\sin\theta_s\sin\theta_y\right)\,,\\ 
\langle e^{-\phi}(0) e^\chi e^{-\phi}(x)\rangle^{\chi\phi}_{C_L}&=-\frac{\pi}{L\sin\theta_x}\,.
\end{split}
\end{equation}
Combining all the sectors, we have
\begin{equation}
\begin{split} 
\langle\varphi,\widehat{\Psi}_{L2}\rangle=C_{\varphi}\left(h-\frac{1}{2}\right)\int_{\frac{1}{2}}^\infty dx\int_{0}^\infty ds\int_{\frac{1}{2}}^\infty &dy\,\frac{e^{1-x-s-y}}{2\pi \sqrt{(x-\frac{1}{2})(y-\frac{1}{2})}} \left(\frac{2\sin\theta_{s}}{L\sin\theta_x\sin\theta_{x+s}}\right)^{h+\frac{1}{2}}\\
&\times\frac{\theta_y\sin^2\theta_x+\theta_x\sin^2\theta_y-\sin\theta_x\sin\theta_s\sin\theta_y}{\sin^{2}\theta_s}\,.
\end{split}
\end{equation}
Let us next consider the first term $\langle\varphi,\widehat{\Psi}_{L1}\rangle$ given by
\begin{equation}
\label{varphi_Psihat1}
\langle\varphi,\widehat{\Psi}_{L1}\rangle=-\int_{\frac{1}{2}}^\infty dx\int_{\frac{1}{2}}^\infty dy\frac{e^{1-x-y}}{\pi \sqrt{(x-\frac{1}{2})(y-\frac{1}{2})}}\langle f \circ\varphi(0)c e^\chi e^{-\phi}(G_{-1/2}\cdot \sigma_L)\sigma_R(x)\rangle_{C_{x+y}}\,.  
\end{equation}
It vanishes for $h\neq\frac{1}{2}$.
For $h=\frac{1}{2}$,
it is evaluated as
\begin{equation}
\langle\varphi,\widehat{\Psi}_{L1}\rangle=C_{\varphi}\, .
\end{equation}
We conclude
that the inner product $\langle \varphi, \widehat{\Psi}_L\rangle$
can be written as a product of the constant $C_{\varphi}$ 
given by (\ref{matter-3-points})
and the universal function $g(h)$ independent of the particular choice of $\varphi_m$ or the bcc operators:
\begin{equation}
\langle\varphi,\widehat{\Psi}_L\rangle
=C_{\varphi} \, g(h) 
\end{equation} 
with
\begin{equation}
\begin{split}
g(h)=\left(h-\frac{1}{2}\right)\int_{\frac{1}{2}}^\infty dx\int_{0}^\infty ds\int_{\frac{1}{2}}^\infty &dy\,\frac{e^{1-x-s-y}}{2\pi \sqrt{(x-\frac{1}{2})(y-\frac{1}{2})}}\left(\frac{2\sin\theta_{s}}{L\sin\theta_x\sin\theta_{x+s}}\right)^{h+\frac{1}{2}} \\
&\times\frac{\theta_y\sin^2\theta_x+\theta_x\sin^2\theta_y-\sin\theta_x\sin\theta_s\sin\theta_y}{\sin^{2}\theta_s}
\end{split}
\end{equation}
for $h\neq\frac{1}{2}$ and $g(\frac{1}{2})=1$.

\end{document}